\documentclass[10pt]{emulateapj}
\usepackage{apjfonts}
\usepackage{graphicx}
\usepackage{psfig}
\newcommand{\begit}{\begin{itemize}}
\newcommand{\enit}{\end{itemize}}
\newcommand{\begen}{\begin{enumerate}}
\newcommand{\enen}{\end{enumerate}}

\newcommand       \be           {\begin{equation}}
\newcommand       \ee           {\end{equation}}
\newcommand       \bea          {\begin{eqnarray}}
\newcommand       \eea          {\end{eqnarray}}

\newcommand{\beqa}{\begin{eqnarray}} 
\newcommand{\eeqa}{\end{eqnarray}}

\begin{document}

\title{Proto-Neutron Star Winds with Magnetic Fields and Rotation}

\author{Brian D. Metzger\altaffilmark{1,3}}\author{Todd A. Thompson\altaffilmark{2,4}}\author{Eliot Quataert\altaffilmark{1}}

\altaffiltext{1}{Astronomy Department and Theoretical Astrophysics Center, 601 Campbell Hall, Berkeley, CA 94720; bmetzger@astro.berkeley.edu, eliot@astro.berkeley.edu }
\altaffiltext{2}{Department of Astrophysical Sciences, Peyton Hall-Ivy Lane, Princeton University, Princeton, NJ 08544; thomp@astro.princeton.edu}
\altaffiltext{3}{Department of Physics, 366 LeConte Hall, University of California, Berkeley, CA 94720}
\altaffiltext{4}{Lyman Spitzer Jr. Fellow}

\begin{abstract}

We solve the one-dimensional neutrino-heated non-relativistic magnetohydrodynamic (MHD) wind problem for conditions that range from slowly rotating (spin period $P\gtrsim10$\,ms) protoneutron stars (PNSs) with surface field strengths typical of radio pulsars ($B\lesssim10^{13}$\,G), to "proto-magnetars" with $B\approx10^{14}-10^{15}$\,G in their hypothesized rapidly rotating initial states ($P\approx1$\,ms).  We use the relativistic axisymmetric simulations of Bucciantini et al.~(2006) to map our split-monopole results onto a more physical dipole geometry and to estimate the spindown of PNSs when their winds are relativistic.  We then quantify the effects of rotation and magnetic fields on the mass loss, energy loss, and thermodynamic structure of PNS winds. The latter is particularly important in assessing PNS winds as the astrophysical site for the $r$-process.  We describe the evolution of PNS winds through the Kelvin-Helmholtz cooling epoch, emphasizing the transition between (1) thermal neutrino-driven, (2) non-relativistic magnetically-dominated, and (3) relativistic magnetically-dominated outflows.  In the last of these stages, the spindown is enhanced relative to the canonical force-free rate because of additional open magnetic flux caused by neutrino-driven mass loss. We find that proto-magnetars with $P\approx 1$ ms and $B \gtrsim 10^{15}$ G drive relativistic winds with luminosities, energies, and Lorentz factors (magnetization $\sigma \sim 0.1-1000$) consistent with those required to produce long duration gamma-ray bursts and hyper-energetic supernovae (SNe). A significant fraction of the rotational energy may be extracted in only a few seconds, sufficiently rapidly to alter the asymptotic energy of the SN remnant, its morphology, and, potentially, its nucleosynthetic yield.  We find that winds from PNSs with somewhat more modest rotation periods ($\approx 2 - 10$ ms) and with magnetar-strength fields produce conditions significantly more favorable for the $r$-process than winds from slowly rotating, non-magnetized PNSs.  Lastly, we argue that energy and momentum deposition by convectively-excited waves may be important in PNS winds.  We show that this further increases the likelihood of successful $r$-process, relatively independent of the PNS rotation rate and magnetic field strength.

\end{abstract}


\section{Introduction}
\label{section:intro}

On timescales $\lesssim$ 1 s following the core collapse of a massive star, neutrino emission from the resulting hot, deleptonizing protoneutron star (PNS) may play an essential role in launching the supernova (SN) shock (Herant et al.~1994; Burrows, Hayes, $\&$ Fryxell 1995; Janka $\&$ M{\"u}ller 1995) or in generating large-scale anisotropies through hydrodynamical instabilities (Blondin et al.~2003; Scheck et al.~2006; Burrows et al.~2006a,b).  Independent of how the explosion is initiated at early times, a small fraction of the PNS's cooling neutrino emission continues heating the surface layers of the PNS, driving a persistent thermal wind into the cavity evacuated by the rapidly-expanding SN shock (Duncan, Shapiro, $\&$ Wasserman 1986; Woosley et al.~1994); this post-explosion neutrino-driven mass loss persists for the duration of the PNS's Kelvin-Helmholtz cooling epoch, which lasts a time $\tau_{{\rm KH}} \sim 10-100$ s (Burrows $\&$ Lattimer 1986; Pons et al.~1999). 

Once the SN shock has been launched and the PNS cooling epoch has begun, neutrino-driven winds from non-rotating non-magnetic PNSs are unlikely to be energetically important on the scale of the kinetic energy of the accompanying SN ($E_{\rm SN}\approx10^{51}$\,ergs).  However, in the presence of a sufficiently strong global magnetic field the dynamics of a PNS's neutrino-heated outflow are significantly altered (e.g., Thompson 2003a,b).  This point is germane because $\sim 10\%$ (Kouveliotou et al.~1994; van Paradijs et al.~1995; Lyne et al.~1998) of young Galactic neutron stars possess significantly stronger surface magnetic fields ($\sim 10^{14}-10^{15}$ G) than those usually inferred from pulsar spin-down estimates (``magnetars''; for a recent review see Woods \& Thompson 2004).  While the precise origin of these large field strengths is uncertain, it has been argued that their amplification occurs via a dynamo during $\tau_{{\rm KH}}$ (Duncan $\&$ Thompson 1992; Thompson \& Duncan 1993).  The efficiency of the dynamo is determined in part by the core's initial rotation rate $\Omega_{0} = 2\pi/P_{0}$, and analytic arguments suggest that the formation of global magnetar-strength fields might require $P_{0} \sim$ 1 ms rotation at birth.  Such rapid rotation would also alter the dynamics of the PNS wind and provide a reservoir of rotational energy significant on the scale of the accompanying SN explosion:
\be 
E_{{\rm rot}} \simeq 2\times 10^{52}\,M_{1.4}\,R_{10}^{2}\,P_{{\rm ms}}^{-2}\,{\rm ergs},
\label{erot}
\ee
where $M_{1.4}$ is the PNS mass in units of 1.4 M$_{\sun}$, $R_{10}$ is the radius of the PNS in units of 10 km, and $P_{{\rm ms}}$ is the initial PNS rotation period in ms.

Previous authors have suggested that if magnetars are indeed born rapidly rotating, their rotational energy could be efficiently extracted through a magnetized wind (Usov 1992; Thompson 1994; Wheeler et al.~2000; Thompson, Chang, \& Quataert 2004, henceforth TCQ).  Indeed, a proto-magnetar wind's energetics, timescale, and potential for highly relativistic outflow resemble those of the central engine required to power long-duration gamma-ray bursts (LGRBs). TCQ argue that an accurate description of magnetized PNS spin-down must include the effects of neutrino-driven mass loss.  The significant mass loss accompanying the Kelvin-Helmholtz epoch may open the otherwise closed magnetosphere into a ``split monopole''-like structure, enhancing the early spin-down rate of the PNS.  In addition, the mass loading of a PNS wind, and hence its potential asymptotic Lorentz factor $\Gamma$, is largely controlled by the neutrino luminosity during the Kelvin-Helmholtz epoch.  This ``baryon-loading'' issue is particularly important in the LGRB context, where $\Gamma \sim 10-1000$ are typically inferred. 

Because magnetar births are relatively frequent, most cannot produce classical LGRBs, which only occur in $\sim 0.1-1\%$ of massive stellar deaths (e.g., Paczynski 2001, Podsiadlowski et al. 2004, Piran 2005); however, a more common observational signature of magnetar birth may include less energetic or mildly relativistic events, which could be observable as X-ray transients or unusual SNe.  Under some circumstances, the asymmetric energy injection from proto-magnetar winds could produce global anisotropies in the SN remnant, as has been detected through polarization measurements of some SNe (e.g., Wang et al.~2001, 2003).  In addition, if significant rotational energy can be extracted sufficiently rapidly following the launch of the SN shock, the nucleosynthetic yield of the SN could be altered (TCQ), which could explain nickel-rich, hyper-energetic SNe such as SN 1998bw (Galama et al.~1998) or SN 2003dh (Hjorth et al.~2003; Stanek et al.~2003).  Lastly, LGRBs occurring without associated SNe (e.g., GRBs 060505, 060614; Fynbo et al.~2006; Gal-Yam et al.~2006; Della Valle et al.~2006) may be accommodated if a proto-magnetar results from the accretion-induced collapse (AIC) of a white dwarf (see \S\ref{section:AIC}).

Quite apart from the possible impact of PNS winds on the surrounding SN shock, PNS winds themselves have often been considered a promising site for the production of $r$-process nuclides.  However, the conditions necessary for a successful third peak ($A \approx 195$) $r$-process have not been realized in detailed studies of non-rotating, non-magnetized PNS winds (e.g., Qian \& Woosley 1996; Cardall \& Fuller 1997; Otsuki et al.~2000; Thompson et al.~2001).  Given the, as yet, unidentified site of Galactic $r$-process enrichment and the relatively large birthrate of magnetars, it is essential to consider what effects a strong magnetic field and rapid rotation might have on nucleosynthesis in PNS winds.  Conversely, if the nucleosynthetic yield from rotating, magnetized PNS winds could be well-determined theoretically, the birth period and magnetic field distribution of neutron stars could perhaps be constrained from observations of heavy elemental abundances.

\subsection{Stages of PNS Evolution}
\label{section:stages}

A sufficiently-magnetized PNS wind goes through at least three distinct stages of evolution following the launch of the SN shock ($t=0$):  

(1) During the earliest phase the PNS surface temperature is so high ($\gtrsim 5$ MeV) and the radius of the PNS is so large ($R_{\nu} \sim 20-50$ km) that, even for a magnetar-strength field, the post-explosion outflow is likely to be purely thermally-driven.  

(2)  By $t \sim$ 1 s the SN shock has propagated to well outside the sonic point of the PNS wind.  The PNS will contract and cool to a point at which, if the surface field is sufficiently strong, the outflow becomes magnetically dominated.  The strong magnetic field enhances angular momentum and rotational energy loss by forcing outgoing fluid elements to effectively corotate with the PNS surface out to the Alfv\'{e}n radius ($R_{A}$) at several stellar radii, in analogy with classic work on non-relativistic stellar winds and the solar wind (e.g., Schatzman 1962, Weber \& Davis 1967, and Mestel 1968).  For sufficiently rapidly rotating PNSs, the spin-down timescale ($\tau_{{\rm J}} \equiv \Omega/\dot{\Omega}$) can be comparable to $\tau_{{\rm KH}}$, implying that much of the rotational energy of the PNS can be extracted in a non-relativistic, but magnetically-dominated wind.  The neutrino luminosity at these relatively early times is still large (e.g., $L_{\nu} \sim 10^{52}$ ergs s$^{-1}$; see Pons et al.~1999, Fig.~14), and, for sufficiently rapid rotation, mass loss is significantly enhanced by centrifugal flinging (TCQ).  The outflow during this phase will be collimated about the PNS rotation axis by magnetic stresses (e.g., Bucciantini et al.~2006; hereafter B06).

(3) As the PNS continues to cool, the luminosity and mass-loading decrease to the point at which corotation is sustained out to nearly the light cylinder ($R_{{\rm L}} \equiv 2\pi c/P \simeq 48\,P_{{\rm ms}}^{-1}$ km).  The spin-down rate then becomes approximately independent of the mass loss rate $\dot{M}$ and the flow becomes relativistic, obtaining high magnetization $\sigma \equiv \Phi_{B}^{2}\Omega^{2}/\dot{M}c^{3}$, where $\Phi_{B}$ is the total open magnetic flux per $4\pi$ steradian (Michel 1969).  Although an ultra-relativistic, pulsar-like wind is inevitable soon after $\tau_{{\rm KH}}$ because $\dot{M}$ abates as the neutrino luminosity vanishes, relativistic outflow can begin prior to the end of the Kelvin-Helmholtz phase.  Because this outflow is accompanied by significant mass loss, it is generally only mildly relativistic.  Although the energy extracted via relativistic winds is primarily concentrated at low latitudes (B06), the confining pressure of the overlying, exploding, stellar progenitor (e.g., Wheeler et.~al 2000; Uzdensky $\&$ MacFadyen 2006) or the walls of the collimated cavity carved by the preceding non-relativistic phase may channel the relativistic outflow into a bipolar, jet-like structure.

The primary focus of this paper is to delineate the magnetic field strengths and rotation rates required to significantly alter the characteristics of early PNS evolution by solving the one-dimensional (1D) non-relativistic MHD, neutrino-heated wind problem for conditions that range from normal pulsars to proto-magnetars.  In particular, we quantitatively explore the transition that occurs between stages (1) and (2) above; with these results we analyze some of the immediate consequences of neutron star birth.  We defer a detailed study of the transition between stages (2) and (3) to future work, but we do explicitly address the parameter space of the $\sigma=1$ boundary.  Relativistic, mass-loaded MHD winds have been studied recently by B06 in two dimensions for both a monopolar and aligned-dipolar field structure, assuming an adiabatic equation of state.  While the work of B06 is critical to understanding the multi-dimensional character of PNS winds (for instance, the degree of collimation and the fraction of open magnetic flux) it does not address the neutrino microphysics necessary for a direct application to PNS environments; this work and that of B06 are thus complementary in studying PNS spin-down in the presence of rapid rotation and a large magnetic field.
\subsection{This Paper} 
In $\S$\ref{section:model} we enumerate the equations of MHD ($\S$\ref{section:equations}), discuss the relevant microphysics ($\S$\ref{section:microphysics}), and elaborate on our numerical methods ($\S$\ref{section:numerical}).  Section \ref{section:results} presents the results of our calculations and examines the regimes of magnetized PNS wind evolution.  Section \ref{section:discussion} discusses the implications of this work, examining the time-evolution of a cooling, magnetized PNS ($\S$\ref{section:evolution}), weighing the implications for LGRBs and hyper-energetic SNe ($\S$\ref{section:LGRB}), and considering the viability for third-peak $r$-process nucleosynthesis in magnetized, rotating PNS winds ($\S$\ref{section:rprocess}).  In $\S$\ref{section:waveheating} we consider the effects that wave heating have on the $r$-process in PNS winds and in $\S\ref{section:AIC}$ we briefly discuss other contexts in which our calculations may be applicable, including the accretion-induced collapse of white dwarfs, ``collapsars,'' and merging neutron star binaries.  Finally, section \ref{section:conclusions} summarizes the conclusions of our work.

\section{PNS Wind Model}
\label{section:model}

\subsection{MHD Equations and Conserved Quantities}
\label{section:equations}

Making the simplifications of Weber $\&$ Davis (1967), we restrict all physical quantities to be solely functions of radius $r$ and time $t$, and confine our analysis to the equatorial plane so that the magnetic field ${\textbf B}$ = [$B_{r},B_{\phi}$] and fluid velocity ${\textbf v}$ = [$v_{r},v_{\phi}$] have no $\theta$ components.  We employ Newtonian gravity for a PNS of mass $M$ acting on gas of density $\rho$.  We also assume that the plasma is a perfect conductor with an isotropic thermal pressure ${\rm P}$.  Under these restrictions the time evolution equations of non-relativistic MHD are 
\be
\frac{\partial\rho}{\partial t} =
-\frac{1}{r^{2}}\frac{\partial}{\partial r}(r^{2}\rho v_{r})
\label{continuity}
\ee
\be 
\frac{\partial v_{r}}{\partial t} = \frac{v_{\phi}^{2}}{r} - v_{r}\frac{\partial v_{r}}{\partial r} - \frac{1}{\rho}\frac{\partial {\rm P}}{\partial r} - \frac{GM}{r^{2}} - \frac{1}{4\pi \rho}\left[\frac{B_{\phi}^{2}}{r} + B_{\phi}\frac{\partial{B_{\phi}}}{\partial r}\right] 
\label{radmomentum}
\ee
\be 
\frac{\partial v_{\phi}}{\partial t} = \frac{1}{r}\left[\frac{B_{r}}{4\pi \rho}\frac{\partial}{\partial r}(r B_{\phi}) - v_{r}\frac{\partial}{\partial r}(rv_{\phi})\right]
\label{phimomentum}
 \ee
\be 
\frac{\partial B_{\phi}}{\partial t} = \frac{1}{r}\frac{\partial}{\partial r}(r[v_{r}B_{\phi} - v_{\phi}B_{r}])
\label{induction}
 \ee
We have neglected neutrino radiation pressure in equation (\ref{radmomentum}) because the neutrino luminosity is always below the neutrino Eddington limit.  In steady state, equation (\ref{continuity}) gives a radially-conserved mass flux:
\be 
\dot{M} = \Delta\Omega r^{2}\rho v_{r},
\label{mdoteq}
\ee
where $\Delta\Omega$ is the opening solid angle of the wind.  While the evolution equations considered are formally valid only in the equatorial plane, quoted values for $\dot{M}$ will be normalized to $\Delta\Omega = 4\pi$, as if the solutions were valid at all latitudes.  As discussed in $\S\ref{section:evolution}$, this normalization is an overestimate because the PNS's closed magnetic flux prevents mass outflow from near the equator and centrifugal flinging concentrates $\dot{M}$ at relatively low latitudes (B06).  

We take the radial magnetic field structure to be that of a ``split monopole'': $B_{r} = B_{\nu}(R_{\nu}/r)^{2}$, where $R_{\nu}$ is the radius of the PNS and $B_{\nu}$ is the monopole surface field strength; this prescription conserves the magnetic flux $4\pi\Phi_{B} \equiv 4\pi r^{2}B_{r}$.  B06 show that the spin-down in dipole simulations can be expressed in terms of an equivalent monopole field, which depends on the fraction of open magnetic flux.  As discussed further in $\S$\ref{section:evolution}, we can therefore relate our monopole spin-down calculations to the more realistic dipole simulations of B06.

In steady state, manipulation of equations (\ref{phimomentum}) and (\ref{induction}) gives the conserved specific angular momentum $\mathcal{L}$ and ``the consequence of induction'' $\mathcal{I}$ (e.g., Lamers $\&$ Cassinelli 1999):
\be 
\mathcal{L} = \mathcal{L}_{{\rm gas}} +  \mathcal{L}_{{\rm mag}} = rv_{\phi} - \frac{rB_{r}B_{\phi}}{4\pi\rho v_{r}} 
\label{ltot}
\ee
\be \mathcal{I} = r(v_{\phi}B_{r}-v_{r}B_{\phi}) 
\label{induct}
\ee
The total rate of angular momentum loss from the PNS is therefore $\dot{J} = \mathcal{L}\dot{M}$.    

For conditions of interest, photons are trapped and advected with the wind, providing no significant energy transport on the timescales of interest (Duncan, Shapiro, $\&$ Wasserman 1986).  Instead, the PNS evolution is controlled by its neutrino luminosity $L_{\nu}$, which provides heating $\dot{q}^{+}_{\nu}$ (per unit mass) above the PNS surface.  Including net neutrino heating ($\dot{q}_{\nu} = \dot{q}^{+}_{\nu} - \dot{q}^{-}_{\nu}$; see $\S$\ref{section:microphysics}) provides a source term in the wind entropy equation:
\be T\frac{dS}{dt} = \dot{q}_{\nu} 
\label{entropyeq},
\ee
where $d/dt \equiv \partial/\partial t + v_{r}(\partial/\partial r)$, $S$ is the wind entropy per unit mass, and $T$ is the wind temperature.  The asymptotic wind entropy $S^{a}$ quantifies the total net heating a parcel of gas experiences as it is carried out by the PNS wind.  As discussed further in $\S$\ref{section:microphysics}, when applying equation (\ref{entropyeq}) we assume that the wind's composition (electron fraction) remains constant.

With net heating the Bernoulli integral $\mathcal{B}$ is not constant with radius (in steady state); instead it receives an integrated contribution from  $\dot{q}_{\nu}$:
\be \dot{M}\Delta\mathcal{B} = \dot{M}\{\mathcal{B}(r) - \mathcal{B}(r_{0})\} = \int_{r_{0}}^{r}\rho \dot{q}_{\nu} \Delta\Omega r'^{2}dr', 
\label{bernoulli}
\ee 
where
\be \mathcal{B} \equiv \frac{1}{2}(v_{r}^{2}+v_{\phi}^{2}) + h - \frac{GM}{r} + \mathcal{L}_{{\rm mag}}\Omega, 
\label{delB}
\ee  
$h \equiv e+{\rm P}/\rho$ is the specific enthalpy, $e$ is the specific internal energy, $\mathcal{L}_{{\rm mag}}\Omega$ is the specific magnetic energy, and $\Omega$ is the stellar rotation rate.  

To assess the impact of the PNS wind on its surroundings, it is useful to define $\eta$, the ratio of the rotational power lost by the PNS, $\dot{E}_{{\rm rot}} = \Omega\dot{J}$, to the asymptotic wind power $\dot{E}^{a} = \mathcal{B}^{a}\dot{M}$:
\be 
\eta \equiv \frac{\Omega \dot{J}}{\dot{E}^{a}} = \frac{\Omega\mathcal{L}}{\mathcal{B}^{a}}, 
\label{eta1}
\ee
 where $\mathcal{B}^{a}$ is the Bernoulli integral evaluated at large radii.  Magnetized winds are typically assumed to have $\eta \approx 1$, but we find that in some cases $\eta \gg 1$ because much of the rotational energy is used to unbind the wind from the PNS (see \S \ref{section:FMR}).

\subsection{Microphysics}
\label{section:microphysics}

The local heating and cooling rates relevant to the Kelvin-Helmholtz cooling phase have been extensively evaluated in efforts to quantify unmagnetized, slowly-rotating PNS winds as an astrophysical site for $r$-process nucleosynthesis (e.g., Qian \& Woosley 1996, Thompson et al.~2001; hereafter T01).  For our calculations we adopt the heating and cooling rates used in the analytic work of Qian \& Woosley (1996, hereafter QW) for $\nu/\bar{\nu}$ annihilation and for the charged-current processes $\nu_{e}+n\leftrightarrow p+e^-$ and $\bar{\nu}_{e}+p\leftrightarrow n+e^+$.  For heating and cooling from inelastic neutrino-lepton scattering we use the rates of T01.  The dominant contributions to the net heating rate are the charged-current processes, but scattering and annihilation become more significant as the entropy of the wind increases and the thermal pressure becomes radiation-dominated.  The scattering and charged-current rates effectively vanish once $T \rightarrow$ 0.5 MeV because the $e^{-}/e^{+}$ pairs annihilate and the nucleons combine into $\alpha$ particles.  We artificially take this cutoff into account by setting $\dot{q}_{\nu} = 0$ for $T < 0.5$ MeV.

Because the charged-current interactions modify the neutron abundance of the outflow the electron fraction $Y_{e}$ should be evolved in addition to the temperature and pressure.  However, the dynamics of the wind are not sensitive to the precise profile of $Y_{e}$ and thus, for simplicity, we take $Y_{e}$ to be fixed at a reasonable asymptotic value at all radii: $Y_{e} = Y_{e}^{a} = 0.4$.  This is a good approximation given how rapidly in radius $Y_{e}^{a}$ obtains in non-rotating, unmagnetized calculations (see T01, Fig.~7) and how relatively weakly the total heating and cooling rates depend on $Y_{e}$ for $Y_{e} \lesssim 0.5$.  

As in T01 we calculate local neutrino fluxes by considering a single (for all neutrino species) sharp, thermal neutrinosphere at $R_{\nu}$.  To the desired accuracy of our calculations this approximation is generally good, even when the density scale height (and hence nucleon-absorption optical depth) is extended by rapid rotation.  For this paper we index stages of the PNS thermal evolution in terms of the anti-electron neutrino luminosity $L_{\bar{\nu}_{e}}$.  We scale all other neutrino luminosities ($L_{\nu_e}, L_{\nu_{\mu}}, L_{\bar{\nu}_{\mu}}, L_{\nu_{\tau}}$, and $L_{\bar{\nu}_{\tau}}$) as in TCQ: $L_{\nu_{e}} = L_{\bar{\nu}_{e}}/1.3 = 1.08 L_{\nu_{\mu}}$, where $\mu$ denotes each of the other four neutrino/anti-neutrino species.  Note that the total neutrino luminosity is then $L_{\nu} \simeq 4.6 L_{\bar{\nu}_{e}}$.  Following T01, all first energy moments at the neutrinosphere ($\langle\epsilon_{\nu}\rangle \equiv \langle E_{\nu}^{2}\rangle/\langle E_{\nu} \rangle$, where $E_{\nu}$ is the neutrino energy) were scaled with luminosity as $\langle\epsilon_{\nu}\rangle \propto L_{\nu}^{1/4}$, anchoring $\{\langle\epsilon_{\nu_{e}}\rangle,\langle\epsilon_{\bar{\nu}_{e}}\rangle,\langle\epsilon_{\nu_{\mu}}\rangle\}$ at $\{11,14,23\}$MeV for $L_{\bar{\nu}_{e},51} = 8,$ where $L_{\bar{\nu}_{e},51}$ is the anti-electron neutrino luminosity in units of $10^{51}$ ergs s$^{-1}$.  Higher energy moments necessary for the heating calculations ($\langle\epsilon^{\,\,n}_{\nu_{e}}\rangle, \langle\epsilon^{\,\,n}_{\bar{\nu}_{e}}\rangle$, etc.) are related to the first through appropriate integrals over the assumed Fermi-Dirac surface distribution.  We should note that the relationship between the neutrino luminosity and mean neutrino energy we have assumed, while a reasonable approximation, is likely to be more complicated.  For example, Pons et al.~(1999) find that the mean energy is roughly constant for the first $\sim$ 10 s of cooling, despite the fact that the neutrino luminosity decreases monotonically (see their Fig.~18).

The neutrino heating and cooling rates discussed above will be modified by the presence of magnetar-strength fields in the heating region due to quantum effects restricting the electron(positron) phase space (Lai $\&$ Qian 1998; Duan $\&$ Qian 2004).  We neglect the
effects that high B have on $\dot{q}_{\nu}$ and defer study of these effects to future work.  In addition, for strong surface magnetic field strengths, heating via the dissipation of convectively-excited MHD waves may become important (Suzuki $\&$ Nagataki 2005).  We assess the importance of wave heating and momentum deposition in $\S$\ref{section:waveheating}.

In this work we include gravitational redshifts, radial Doppler shifts, and modifications to the ``effective solid angle'' (and hence local neutrino flux) presented by the neutrinosphere in the curved spacetime.  The latter effect is described in Salmonson $\&$ Wilson (1999), while the Doppler and redshifts can be combined into the simple, approximate prescription relevant for all neutrino species:   
\be 
\langle \epsilon_{\nu}^{\,\,n} \rangle = (\phi_{Z}\phi_{D})^{n+3} \langle \epsilon_{\nu}^{\,\,n}\rangle |_{r=R_{\nu}}, 
\ee
where
\be 
\phi_{Z} \equiv \alpha(R_{\nu})/\alpha(r)\,\,,\,\,\phi_{D} \equiv \gamma(1 - v_{r}/c), 
\ee
$\alpha(r) \equiv \sqrt{1-(2GM/c^{2}r)}$, $\gamma^{-1} \equiv \sqrt{1-(v_{r}^{2}+v_{\phi}^{2})/c^{2}}$, and we have assumed that at radii where the equatorial flow becomes mildly relativistic, typical neutrinos will primarily be moving radially.  We emphasize that while we include neutrino gravitational redshifts in calculating heating rates we calculate the wind dynamics in Newtonian gravity.  Including the effects of the deeper general-relativistic (GR) potential of a Schwarzschild metric lowers $\dot{M}$ and increases $S^{a}$, $Y_{e}^{a}$, and the asymptotic wind speed $v^{a}$ (Fuller $\&$ Qian 1996; Cardall $\&$ Fuller 1997).

Our model's equation of state (EOS) includes contributions from photon radiation, ideal nucleons, and relativistic, degenerate electrons and positrons.  Non-relativistic nucleons generally dominate the EOS near the PNS surface, but within a density scale height above $R_{\nu}$ the flow becomes radiation-dominated.  As with the heating/cooling rates, high magnetic field effects on the EOS are ignored.

\subsection{Numerical Method}
\label{section:numerical}

In the steady-state Weber-Davis wind, three critical points occur in the radial momentum equation at radii where the outflow velocity matches the local phase speed of infinitesimal fluid disturbances.  The steady-state eigenvalues $\dot{M}, \mathcal{L},$ and $\mathcal{B}$ (eqs.~[\ref{mdoteq}], [\ref{ltot}], and [\ref{delB}]) are fixed by the requirement that the solution pass smoothly through the slow-magnetosonic, Alfv\'{e}nic, and fast-magnetosonic points.  Physically, we choose these solutions over sub-magnetosonic ``breezes'' because we assume the SN shock or fallback pressures at large radii are insufficient to stifle the strong ram-pressures of the wind.  However, fallback at early times is not well-understood because it depends sensitively on the mechanism for launching the SN shock (Chevalier 1989; Woosley $\&$ Weaver 1995) and thus this issue deserves further attention.  Although the PNS radius, rotation rate, and neutrino luminosity evolve in time, for realistic wind conditions the timescale required for any of the MHD wavemodes to traverse all critical points is always much shorter than the timescale over which the wind characteristics appreciably change (e.g., $\tau_{{\rm KH}}$ or the spin-down timescale $\tau_{{\rm J}}$).  For this reason a time-series of steady-state solutions is generally sufficient to accurately model the wind during all phases of the PNS evolution.  However, precisely because all physical solutions must pass through each critical point, in the time-independent formulation of this problem boundary conditions must be placed on the wind solution at these locations.  To avoid this numerically complicated singularity structure, we have instead solved the more complete, time-dependent version of the problem using the 6th order space/3rd order time, ``inhomogeneous'' 2N-RK3 scheme of Brandenburg (2001).

Our code evolves the variables ($\rho,T,v_{r},v_{\phi},B_{\phi}$).  The value of the PNS mass $M$, neutrinosphere radius $R_{\nu}$, magnetic flux $ \Phi_{B} = B_{\nu}R_{\nu}^{2}$, stellar rotation rate $\Omega$, and neutrino luminosity $L_{\nu}$ are the parameters that uniquely identify a wind solution.  We use $M$ = 1.4 M$_{\sun}$ and $R_{\nu} = 10$ km in all of our calculations.  Because our code is intrinsically non-conservative, we use the constancy of $\dot{M}$, $\mathcal{L}$, and $\mathcal{I}$ and equation (\ref{bernoulli})'s constraint on $\mathcal{B}$ as independent checks on the code's numerical accuracy.  For numerical stability, an artificial viscosity of the form $\nu \nabla^{2}$ is included in the evolution of each variable, where $\nu$ is an appropriately-scaled kinematic viscosity (e.g., Brandenburg 2001).

We chose the location of the outer boundary, generally at $r \approx 1000$ km, as a compromise between the run-time to reach steady-state and the desire to minimize the effects of artificially forcing the fast point on the computational grid (see discussion in $\S$\ref{section:bcs}).  We space the radial grid logarithmically, choosing the number of grid points (generally $500-2000$) to obtain the desired level of conservation of $\dot{M}, \mathcal{L}, \mathcal{I},$ and $\mathcal{B}$ while simultaneously maintaining large enough artificial viscosity to maintain code stability.  With sufficient resolution and low enough viscosity the code shows radial conservation of all eigenvalues to $\lesssim 1\%$ across the entire grid, although we did not require this level of conservation for all solutions so that we could efficiently explore the parameter space of wind properties.  The mass loss rate was the most difficult eigenvalue to conserve yet is accurate (relative to its fully converged value) to at least $\sim 10\%$ for all solutions presented in this paper.

\subsubsection{Boundary Conditions}
\label{section:bcs}

The azimuthal speed $v_{\phi}$ at $R_{\nu}$ is set to enforce $v_{\phi,\nu} = R_{\nu}\Omega + v_{r,\nu}B_{\phi,\nu}/B_{\nu}$, where a subscript $\nu$ denotes evaluation at $R_{\nu}$ and where $\Omega$ is the stellar rotation rate that defines the angular speed of the rotating frame in which the surface electric field vanishes (MacGregor $\&$ Pizzo 1983).  In all cases we consider, $v_{\phi,\nu} \simeq R_{\nu}\Omega$.  The temperature at the PNS surface $T(R_{\nu})$ (generally $\approx$ 5 MeV) is set by requiring that the net heating rate $\dot{q}_{\nu}$ vanish; this assumes the inner atmosphere is in LTE (Burrows $\&$ Mazurek 1982).  Given our assumption that matter at the inner grid point maintains LTE, we fix the density at $R_{\nu}$ to be $\simeq 10^{12}$ g cm$^{-3}$ so that the neutrino optical depth $\tau_{\nu}$ at the PNS surface reaches $\sim$ 2/3, thereby defining a neutrinosphere.  For slow rotation we find that the solution outside the inner few scale heights remains relatively insensitive to our choice for the inner density, although we find that $\dot{M}$ depends somewhat sensitively on $\rho(R_{\nu})$ as the PNS rotation rate increases approaching break-up.  

If all three critical points are captured on the numerical grid, the outer boundary conditions are not in causal contact with the interior wind and will have no effect on its steady-state eigenvalues.  However, as the temperature of the wind declines, the fast magnetosonic point moves to very large radii; in fact, as the sound speed $c_{{\rm s}} \rightarrow 0$ the fast point formally approaches infinity (Michel 1969).  For this reason, the fast point is difficult to keep on the computational grid.  Solutions without the fast point captured on the grid are sensitive to the outer boundary condition, with different choices altering, for instance, the spin-down rate.  Therefore, to artificially force the fast point on the grid we increase the outer radial velocity boundary-condition until the fast point is captured.  Otherwise equivalent solutions with the fast point naturally located and artificially placed on the grid were compared in several cases; we found that although the velocity structure changes at radii far outside the Alf\'{v}en radius, our imposed boundary condition had little effect on the eigenvalues of the problem and the correct asymptotic speed was obtained (albeit prematurely in radius).  Since the eigenvalues uniquely determine the steady-state solution, this technique, when necessary, was a useful expedient to obtain the desired transmagnetosonic solution.

\begin{figure*}
\centerline{\hbox{\psfig{file=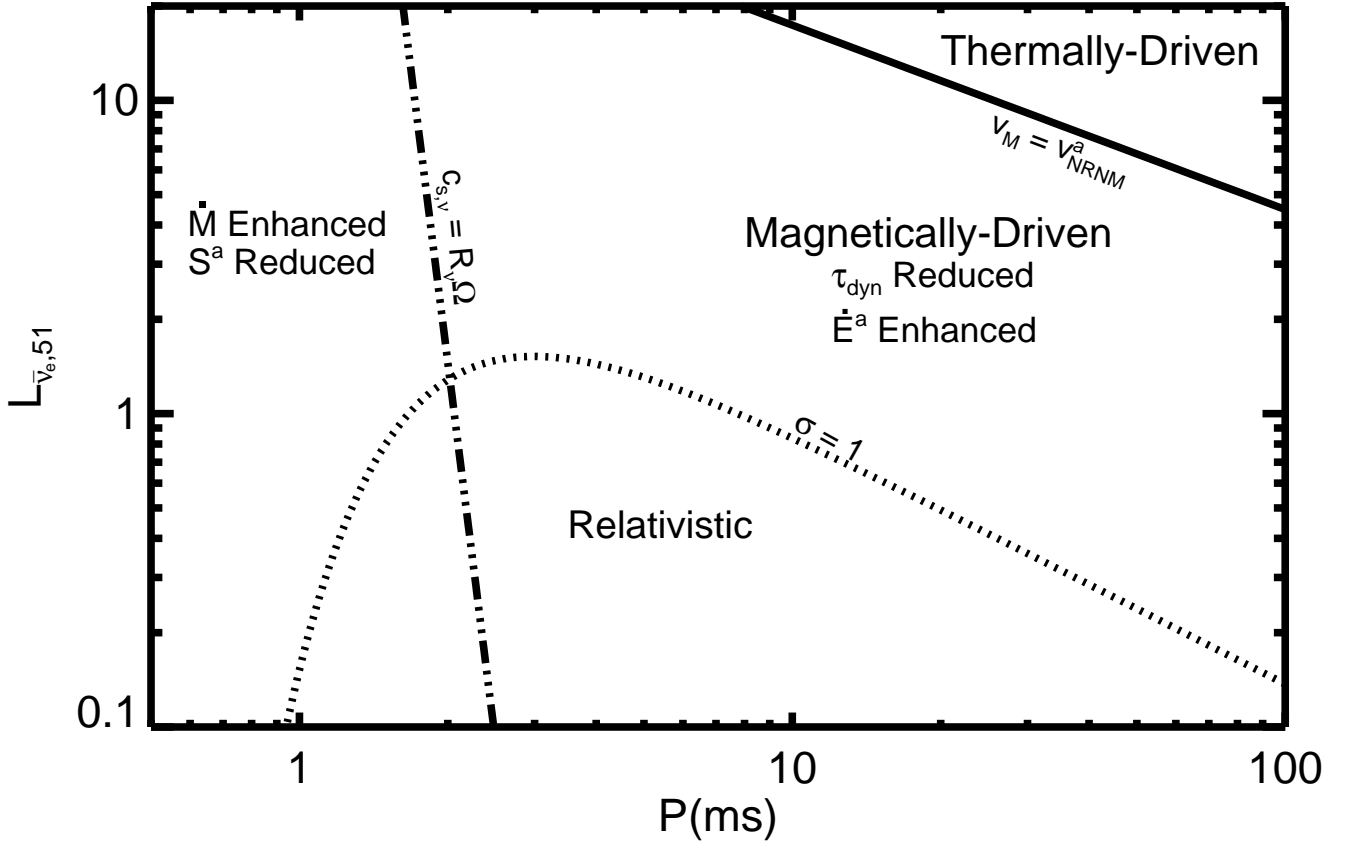,width=18cm}}}
\figcaption[x]{ The regimes of PNS winds in the space of $L_{\bar{\nu}_{e}} (= L_{\bar{\nu}_{e},51}\times 10^{51}$ ergs s$^{-1}$) and rotation period $P$ for a representative strongly magnetized PNS: $B_{\nu} = 2.5\times 10^{14}$ G, $R_{\nu} = 10$ km, and $M = 1.4 $ M$_{\sun}$.  A monopole field geometry is assumed.  A cooling PNS will evolve from high $L_{\bar{\nu}_{e}}$ to low $L_{\bar{\nu}_{e}}$ in this diagram, with $L_{\bar{\nu}_{e},51} \sim 10$ at a time $t_{0} \sim$ 1 s following the launch of the SN shock to $L_{\bar{\nu}_{e},51} \sim 0.1$ by the end of the Kelvin-Helmholtz phase ($\tau_{{\rm KH}} \sim 10-100$\,s after explosion).  The solid line (eq.~[\ref{omegac}]) shows the boundary between PNS winds that are primarily thermally-driven by neutrino heating (high $L_{\nu}$; slow rotation) and winds that are primarily magneto-centrifugally driven (low $L_{\nu}$; rapid rotation); for the latter, the wind power $\dot{E}^{a}$ is enhanced (see eq.~[\ref{edotfmr}]) and the dynamical time $\tau_{{\rm dyn}}$ is reduced (see Fig.~\ref{plot:tdynplot}).  The dotted line ($\sigma = 1$; eq.~[\ref{sigma}]) shows the boundary between non-relativistic and relativistic magnetically-driven winds.  For sufficiently rapid rotation ($P\lesssim 2-3$ ms; the dot-dashed line), the mass-loss from the PNS ($\dot{M}$) is enhanced because of centrifugal flinging (eq.~[\ref{mdotcent}]) and the asymptotic wind entropy $S^{a}$ is reduced (eq.~[\ref{sa}]) because matter moves more rapidly through the region of significant neutrino heating.  The dynamical time and entropy are important for nucleosynthesis in the wind ($\S\ref{section:rprocess}$).  For $B_{\nu} \gtrsim 2.5\times 10^{14}$ G the thermally-driven region shrinks (to longer $P$ and higher $L_{\bar{\nu}_{e}}$; eq.~[\ref{omegac}]) and the relativistic region expands (to higher $L_{\bar{\nu}_{e}}$; eq.~[\ref{sigma}]).  For $R_{\nu}\gtrsim$ 10 km, as will occur at early times when the PNS is still contracting, the wind is likely to be thermally driven.  This figure illustrates the wide range of conditions under which PNS winds will be magnetically-driven, although it should be cautioned that the surface dipole field $B_{\nu}^{{\rm dip}}$ associated with the monopole field $B_{\nu}$ scales as $B_{\nu}^{{\rm dip}} \propto B_{\nu}P$ (see eq.~[\ref{bmoneff}]), which means that, for large $P$, the true dipole field appropriate to this diagram is much greater than the monopole value of $B_{\nu} = 2.5\times 10^{14}$ G. \\
\label{plot:regimes}}
\end{figure*}

\section{Results}
\label{section:results}

Figure \ref{plot:regimes} summarizes the physical regimes of PNS winds as a function of $L_{\bar{\nu}_{e}}$ and rotational period $P$ for a representative strongly magnetized PNS: $B_{\nu} = 2.5\times 10^{14}$ G, $M = 1.4$ M$_{\sun}$, and $R_{\nu} = 10$ km.  A cooling PNS of fixed surface field will traverse a path from high to low $L_{\nu}$ in this diagram, reaching $L_{\bar{\nu}_{e},51} \sim 0.1$ at $t = \tau_{{\rm KH}}$ (see eq.~[\ref{lumevo}]).  If the spindown timescale $\tau_{{\rm J}}$ is less than $\tau_{{\rm KH}}$, the PNS evolves to higher $P$ during $\tau_{{\rm KH}}$, but otherwise, the PNS evolves from higher to lower $L_{\nu}$ at roughly constant $P$ (see Tables \ref{table:solutiontable} and \ref{table:solutiontable2} for representative $\tau_{{\rm J}}$).  The regions in Figure \ref{plot:regimes} correspond to the different  wind phases outlined in $\S$\ref{section:stages}: (1) a thermally-driven wind at high $L_{\nu}$ and long $P$; (2) a non-relativistic, magnetically-driven wind at high $L_{\nu}$ and short $P$; and (3) a relativistic, magnetically-driven wind at low $L_{\nu}$ and short $P$.  In addition to these different regimes, Figure \ref{plot:regimes} illustrates the range of rotation periods for which $\dot{M}$ is enhanced by centrifugal flinging (roughly $P\lesssim 2-3$ ms; see also Fig.~\ref{plot:mdotomega}) and for which $\tau_{{\rm dyn}}$, the dynamical time at $T=0.5$ MeV (eq.~[\ref{taudef}]), and the asymptotic wind entropy $S^{a}$ are altered from their non-rotating, non-magnetized values (which has important consequences for $r$-process nucleosynthesis; see $\S$\ref{section:rprocess}).  In this section, we present and discuss the properties of solutions for a range of parameters ($B_{\nu}, P, L_{\bar{\nu}_{e}}$) that span each of the non-relativistic regimes illustrated in Figure \ref{plot:regimes}.  Some of the properties of wind solutions at $L_{\bar{\nu}_{e}} = 8\times 10^{51}$ ergs s$^{-1}$, which corresponds to a relatively early stage in the PNS cooling evolution, are given in Table \ref{table:solutiontable}.  Table \ref{table:solutiontable2} compares the properties of wind solutions with $B_{\nu} = 2.5\times 10^{15}$ G at two different neutrino luminosities ($L_{\bar{\nu}_{e}} =  8\times 10^{51}$ and $3.5\times 10^{51}$ ergs s$^{-1}$).

\subsection{Thermally-Driven Winds}

\begin{figure}[t]
\centerline{\hbox{\psfig{file=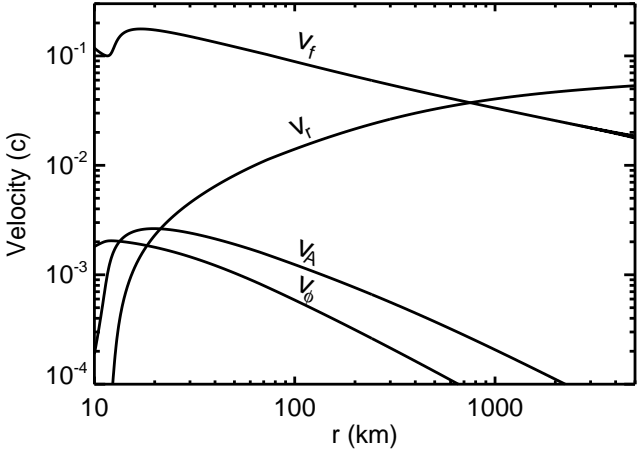,width=9.5cm}}}
\figcaption[x]{Velocity profile for a thermally-driven wind with $L_{\bar{\nu}_{e},51} = 8$, $B_{\nu} = 10^{13}$ G, and $\Omega = 50$  s$^{-1}$ ($P \approx 130$ ms).  The variables $v_{{\rm r}}$, $v_{\phi}$, $v_{{\rm A}}$, and $v_{{\rm f}}$ are the radial, azimuthal, Alfv\'{e}n, and fast magnetosonic speeds, respectively; the fast(Alfv\'{e}n) speed is also approximately the adiabatic sound(slow) speed for thermally-driven winds.  This solution has $\dot{M} \approx 1.4\times 10^{-4} $M$_{\sun}$ s$^{-1}$, $\sigma \approx 3 \times 10^{-8}$, $\dot{E}^{a} \simeq 4\times 10^{47}$ ergs s$^{-1}$, and $\tau_{{\rm J}} \sim 880$ s (see Table \ref{table:solutiontable}).  The Michel speed for this solution is $v_{M}=\sigma^{1/3}c \simeq 0.003$ c, which is less than the thermally-driven asymptotic speed actually obtained ($\approx 0.06$ c); hence, the magnetic field and rotation rate have no significant effect on the acceleration of the wind.
\label{plot:thermal}}
\end{figure}

The thermally-driven region in Figure \ref{plot:regimes} corresponds to conditions under which the PNS outflow is driven primarily by neutrino heating; the magnetic field and the rotation rate are unimportant in either accelerating or setting the mass loss rate of the wind.  Figure \ref{plot:thermal} shows the velocity structure of such an effectively non-rotating, non-magnetized (NRNM) solution for $L_{\bar{\nu}_{e},51} = 8$, $B_{\nu} = 10^{13}$ G, and $\Omega = 50$ s$^{-1}$ ($P\simeq$ 126 ms).   Notice that the Alfv\'{e}n radius is relatively close to the PNS surface ($R_{A} \approx$ 20 km) and that the sonic point (corresponding to the fast point in the NRNM limit) is at a much larger radius ($R_{{\rm s}} \approx$ 750 km, approximately the ``Parker radius'', $R_{{\rm p}} \simeq GM/(v^{a})^{2}$, of an equivalent polytropic wind, where $v^{a}$ is the asymptotic wind speed).  Although the magnetic field and rotation rate are low enough that they have no effect on the wind energetics, the Alfv\`{e}n radius $R_{A}$ is still above the PNS surface.  Angular momentum loss is thus enhanced by a factor of $(R_{A}/R_{\nu})^{2} \approx 4$ over an unmagnetized wind.  In this respect, PNS winds such as that shown in Figure \ref{plot:thermal} are analogous to the solar wind (which is also primarily thermally-driven, but has $R_{A} \gtrsim R_{\odot}$).

\begin{figure}[t]
\centerline{\hbox{\psfig{file=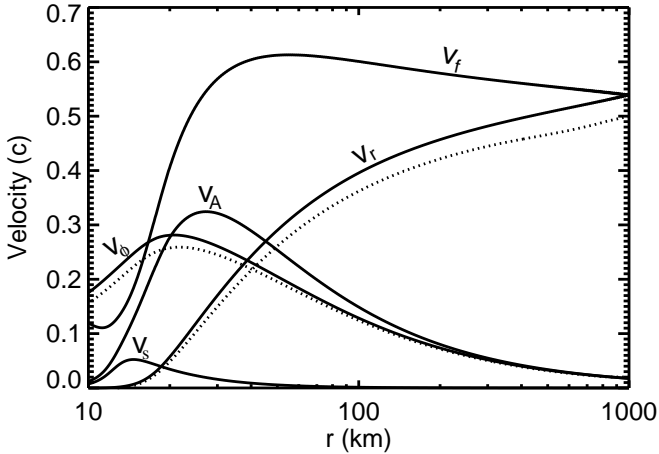,width=9.5cm}}}
\figcaption[x]{Velocity profile for a magneto-centrifugally-driven wind with $L_{\bar{\nu}_{e},51} = 8$, $B_{\nu} = 10^{15}$ G, and $\Omega = 5000$  s$^{-1}$ ($P \simeq 1.26$ ms); the variables $v_{{\rm r}}$, $v_{\phi}$, $v_{{\rm A}}$, $v_{{\rm f}}$, and $v_{{\rm s}}$ are the radial, azimuthal, Alfv\'{e}n, fast, and slow magnetosonic speeds, respectively.  This solution has $\dot{M} \simeq 3.0 \times 10^{-3} $M$_{\sun}$ s$^{-1}$, $\sigma \simeq 0.16$, $\tau_{{\rm J}} \simeq 9$ s, and $\dot{E}^{a} \simeq 2.3\times 10^{51}$ ergs s$^{-1}$.  Comparison plots of $v_{r}$ and $v_{\phi}$ (dashed lines) are for a $\gamma = 1.15$ polytropic wind with similar $\dot{M}$, $\Omega$, $B_{\nu}$, and inner temperature.  Notice that the radius of the slow point (approximately the sonic point; where $v_{r} = v_{{\rm s}}$) is very close to the value expected in the magneto-centrifugal limit: $R_{{\rm s,cf}} \approx$ 19.6 km (eq.~[\ref{rcf}]).
\label{plot:fmr}}
\end{figure}

Many of the relevant results for NRNM PNS winds (such as Fig.~$\ref{plot:thermal}$) are approximated analytically and verified numerically in QW.  QW show that $\dot{M} \propto L_{\bar{\nu}_{e}}^{5/3}\langle\epsilon_{\bar{\nu}_{e}}\rangle^{10/3}$ (their eqs.~[58a,b]); hence, if $L_{\nu} \propto \langle\epsilon_{\nu}\rangle^{4}$, as we have assumed, QW find that the neutrino-driven mass loss rate is approximately given by $\dot{M}_{{\rm NRNM}}$ $ \simeq 3\times 10^{-4}(L_{\bar{\nu}_{e},51}/8)^{2.5}$M$_{\sun}$ s$^{-1}$, where $L_{\bar{\nu}_{e}} = L_{\bar{\nu}_{e},51}\times 10^{51}$ ergs s$^{-1}$.  Our calculations find that $\dot{M}_{{\rm NRNM}} \propto L_{\nu}^{2.5}$ as well, but with a normalization lower than that of QW:
\be 
\dot{M}_{{\rm NRNM}} \simeq 1.4\times 10^{-4}(L_{\bar{\nu}_{e},51}/8)^{2.5}\,{\rm M}_{\sun}\,{\rm s}^{-1}, 
\label{mdotnrnm}
\ee 
primarily because we have included neutrino redshifts in our heating rates.  Somewhat coincidentally, T01 found a result similar to equation (\ref{mdotnrnm}) from calculations incorporating GR.

Because neutrino heating is so concentrated near the PNS surface, NRNM winds are barely unbound in comparison to the PNS escape speed (non-relativistically, $v_{{\rm esc}}(R_{\nu}) \approx 0.64$ c); indeed, from Figure ~\ref{plot:thermal} we find an asymptotic speed $v_{{\rm NRNM}}^{a} \approx 0.06$ c at $L_{\bar{\nu}_{e},51} = 8$.  T01 found that $v_{{\rm NRNM}}^{a}$ $ \simeq 0.1$ c $(L_{\bar{\nu}_{e},51}/8)^{0.3}$ in GR at high $L_{\nu}$.  Although we find that $v^{a}_{{\rm NRNM}}$ scales in approximately the same way with $L_{\bar{\nu}_{e}}$, our asymptotic speeds are lower than those obtained by T01 primarily because we have used a more shallow, Newtonian gravitational potential.  In NRNM winds the asymptotic wind power is entirely gas kinetic energy: $\dot{E}_{{\rm NRNM}}^{a} \simeq (1/2)\dot{M}_{{\rm NRNM}}(v^{a}_{{\rm NRNM}})^{2}$; hence, from our results for $\dot{M}_{{\rm NRNM}}$ and $v^{a}_{{\rm NRNM}}$ we find
\be 
\dot{E}_{{\rm NRNM}}^{a} \simeq 4\times 10^{47}(L_{\bar{\nu}_{e},51}/8)^{3.2}\,{\rm ergs}\,\,{\rm s}^{-1}.
\label{edotnrnm}
\ee
Since the time spent at $L_{\bar{\nu}_{e},51} \sim 8$ is only $\sim 1$\,s (see $\S$\ref{section:evolution}), the total energy extracted during the Kelvin-Helmholtz epoch in a NRNM wind is $\sim 10^{47}-10^{48}$ ergs.

\subsection{Magnetically-Driven Winds}
\label{section:FMR}

For a PNS with a given neutrino luminosity, larger rotation rates and magnetic field strengths lead to additional acceleration in the outer, supersonic portions of the wind.  If the rotation rate and magnetic field are high enough, they will dominate the wind acceleration at large radii.  This is the ``Fast Magnetic Rotator'' (FMR) limit, using the stellar-wind terminology of Belcher $\&$ MacGregor (1976); see also Lamers $\&$ Cassinelli (1999).  An approximate criteria for this limit is that the magnetically-driven asymptotic speed, given roughly by the Michel speed $v_{M} \equiv (B_{\nu}^{2}R_{\nu}^{4}\Omega^{2}/\dot{M})^{1/3} = \sigma^{1/3}c$ (Michel 1969), exceeds the asymptotic speed obtained if the wind were entirely thermally-driven by neutrino heating ($v_{{\rm NRNM}}^{a}$).  Using our result for $v_{{\rm NRNM}}^{a}$, the wind will be in the FMR limit for magnetizations 
\be 
\sigma \gtrsim 2\times 10^{-4}(L_{\bar{\nu}_{e},51}/8)^{0.9} \equiv \sigma_{{\rm FMR}}.
\ee
Using equation (\ref{mdotnrnm}) to relate $L_{\bar{\nu}_{e}}$ and $\dot{M}$, our calculations imply that the PNS wind is magnetically-driven below the critical rotation period 
\be 
P_{{\rm FMR}} \simeq 15\,(L_{\bar{\nu}_{e},51}/8)^{-1.7}B_{14}\,{\rm ms}, 
\label{omegac}
\ee
where $B_{\nu} = B_{n}\times 10^{n}$ G.  The $P = P_{{\rm FMR}}$ boundary in Figure \ref{plot:regimes} is marked by a solid line.  The magnetically-driven regime encompasses a large range of PNS parameter space and hence generically describes most of a strongly magnetized PNS's evolution.  In contrast, the wind from a relatively weakly magnetized PNS will only be dominated by magneto-centrifugal forces late in the cooling epoch.  The conditions necessary for the magnetically-driven phase to dominate the total energy and mass loss during the Kelvin-Helmholtz phase will be discussed further in $\S\ref{section:evolution}$.

Figure \ref{plot:fmr} shows the velocity structure of a magnetically-driven wind from a PNS with $L_{\bar{\nu}_{e},51} = 8$, $B_{\nu} = 10^{15}$ G and $\Omega = 5000$ s$^{-1}$ ($P \simeq 1.3$ ms).  As the profile of $v_{\phi}$ in Figure \ref{plot:fmr} indicates, the wind corotates to $\approx 25$ km, which is far inside $R_{A} \approx 46$ km because the magnetic field carries a significant fraction of the angular momentum.  In addition, because the wind is magnetically-driven, the wind speed at large radii is almost an order of magnitude larger than in a NRNM wind: $v_{r} = 0.54$ c $\approx v_{M}$ obtains at the outer grid point.  The sonic point of the wind (corresponding to the slow point in the FMR limit) is now inside $R_{A}$, less than one stellar radius off the surface; this is expected because analytic considerations show that with increasing $\Omega$, the location of the sonic radius $R_{{\rm s}}$ decreases from a value of order the Parker radius to a value independent of the local thermodynamics (see Lamers $\&$ Cassinelli 1999)\footnote{The subscript ``cf'', here and below, stands for ``centrifugal'' and relates to the limit described by equation (\ref{omegacf}).}:
\be 
R_{{\rm s},{\rm cf}} \equiv \left({GM}/{\Omega^{2}}\right)^{1/3} \simeq 17  P_{{\rm ms}}^{2/3}\,{\rm km}.
\label{rcf}
\ee
For comparison with our solution, Figure \ref{plot:fmr} shows the velocity structure of an adiabatic wind ($\gamma = 1.15$) with approximately the same $\dot{M}$, $\Omega$, $B_{\nu}$, and surface temperature as our neutrino-heated wind.  The adiabatic solution agrees well with the neutrino-heated solution because, although $\dot{M}$ is primarily set by $L_{\nu}$, once $\dot{M}$ is specified the velocity structure of the magnetically-driven wind becomes relatively independent of the details of the neutrino microphysics.  This agreement implies that we can accurately map our 1D neutrino-heated calculations onto multi-dimensional polytropic calculations that employ similar boundary conditions and a similar effective adiabatic index (see $\S$\ref{section:evolution}).

For FMR winds, the Michel speed obtains at large radii and the asymptotic wind power is therefore enhanced relative to equation (\ref{edotnrnm}):
\begin{eqnarray} \dot{E}^{a}_{{\rm FMR}} = \dot{M}\mathcal{B}^{a} \simeq (3/2)\dot{M}v_{M}^{2} \simeq \nonumber \\
10^{50}\,B_{14}^{4/3}\dot{M}_{-3}^{1/3}P_{{\rm ms}}^{-4/3}\,{\rm ergs}\,\,{\rm s}^{-1}, 
\label{edotfmr}
\end{eqnarray}
where $\dot{M} = \dot{M}_{-3}\times 10^{-3} $ M$_{\sun}$s$^{-1}$ and $\mathcal{B}^{a} = \dot{E}^{a}/\dot{M} = (v_{M}^{3}/v^{a}) + (v^{a})^{2}/2 \simeq (3/2)v_{M}^{2}$ is the Bernoulli integral at large radii in the FMR limit, and is composed of 2/3 magnetic and 1/3 kinetic energy.  

\begin{figure}
\centerline{\hbox{\psfig{file=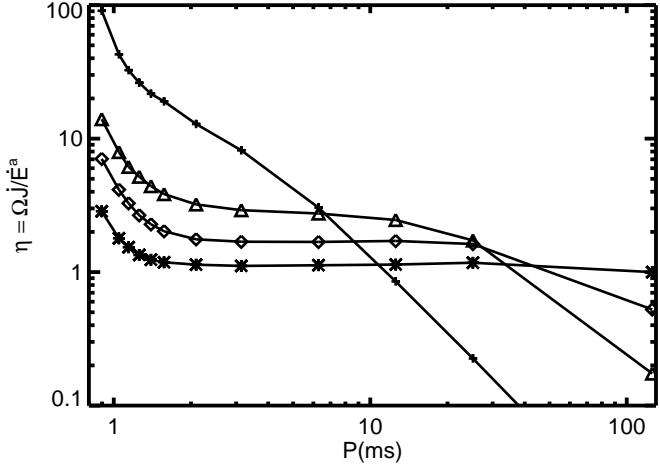,width=9.cm}}}
\figcaption[x]{$\eta = \Omega\dot{J}/\dot{E}^{a}$ is the ratio of spin-down power lost by the PNS to the asymptotic wind power (eq.~[\ref{eta1}]).  This figure shows $\eta$ as a function of rotation period $P$ for $L_{\bar{\nu}_{e}} = 8\times 10^{51}$ ergs s$^{-1}$ and monopole magnetic field strengths $B_{\nu}= 10^{13}$ G (cross), $10^{14}$ G (triangle), $2.5\times 10^{14}$ G (diamond), and $10^{15}$ G (asterisk).  For the highly magnetized solutions (with $R_{A}$ well off the surface) nearly all of the extracted rotational energy escapes to large radii ($\eta \simeq 1$), but for low $B_{\nu}$ and high $\Omega$ most of the rotational energy is used to unbind the wind and hence $\eta \gtrsim 1$.  For slowly rotating, thermally-driven winds (low $B_{\nu}$ and $\Omega$) $\eta \ll 1$ because the rotational power lost by the PNS is insignificant in comparison to the thermal energy supplied by neutrino heating.
\label{plot:eta}}
\end{figure}

To calculate the angular momentum lost by the PNS, we note that for any super-Alfv\'{e}nic outflow, equation (\ref{ltot}), equation (\ref{induct}), and conservation of magnetic flux require that the conserved specific angular momentum obey $\mathcal{L} = \Omega R_{A}^{2},$ where $R_{A}$ is defined by the position where the radial outflow speed matches the radial Alfv\'{e}n speed: $v_{r}(R_{A}) = B_{r}(R_{A})/\sqrt{4\pi \rho(R_{A})} \equiv v_{A}$.  We estimate the location of the Alfv\'{e}n point in terms of $\eta = \Omega\dot{J}/\dot{E}^{a} =  \Omega^{2}R_{A}^{2}/\mathcal{B}^{a}$ defined in equation (\ref{eta1}):
\be 
R_{A}^{2}\Omega^{2} \simeq (3/2)\eta v_{M}^{2}, 
\label{ravm}
\ee
so that\footnote{TCQ took $\Omega R_{A} = v_{r}(R_{A}) = v_{M}$ and assumed $\eta = 1$.  We find that in some circumstances the $\eta=1$ assumption is not applicable, even in the FMR limit (see Fig.~\ref{plot:eta}).} \be R_{A}|_{{\rm FMR}} \simeq 11\,\eta^{1/2}\,B_{14}^{2/3}\dot{M}_{-3}^{-1/3}P_{{\rm ms}}^{1/3}\,{\rm km}.
\label{ra}
\ee

Figure \ref{plot:eta} shows $\eta$ for our wind solutions with $L_{\bar{\nu}_{e},51} = 8$ for several surface magnetic field strengths (see also Table \ref{table:solutiontable}).  Recall that $1/\eta$ represents the fraction of the extracted rotational energy from the PNS available to energize the surrounding environment.  In the limit that $v_{M} \gtrsim v_{{\rm esc}}$, we find that $\eta \approx 1$ and almost all of the rotational energy lost by the PNS emerges as asymptotic wind power; this limit is normally assumed when considering magnetized stellar spin-down.  However, as the low $B_{\nu}$ solutions in Figure \ref{plot:eta} illustrate, winds with short rotation periods and $v_{{\rm NRNM}}^{a} < v_{M} < v_{{\rm esc}}$ can be magnetically-driven and yet have $\eta \gg 1$.  The primary reason for this is that at high $\Omega$ the neutrino heating rate per unit mass is significantly reduced below its NRNM value because centrifugally-accelerated matter spends less time in the region where neutrino heating is important.  Because the wind absorbs less of the neutrino energy, the magnetic field becomes more important for unbinding the matter from the PNS.  Consequently, only a fraction of the rotational energy extracted at the PNS surface reaches large radii.  

In the limit of thermally-driven solutions with very low $B_{\nu}$ (even lower than in Fig.~\ref{plot:thermal}, such that the Alfv\'{e}n radius is interior to the stellar surface), $R_{\nu}$ is the lever arm for angular momentum loss; thus $\mathcal{L} = \Omega R_{\nu}^{2}$ and hence $\eta = \Omega^{2}R_{\nu}^{2}\dot{M}_{{\rm NRNM}}/\dot{E}^{a}_{{\rm NRNM}} \approx 24(L_{\bar{\nu}_{e},51}/8)^{-0.6}P_{{\rm ms}}^{-2}$ (using eqs.~[\ref{mdotnrnm}] and [\ref{edotnrnm}]).  For $P \gg 1$ ms thermally-driven winds have $\eta \ll 1$, which explains why $\eta$ decreases rapidly for solutions with large $P$ in Figure \ref{plot:eta}.  Physically, this is because for slow rotation rates the rotational power lost by the PNS is insignificant in comparison to the thermal energy supplied by neutrino heating.

\begin{figure}
\centerline{\hbox{\psfig{file=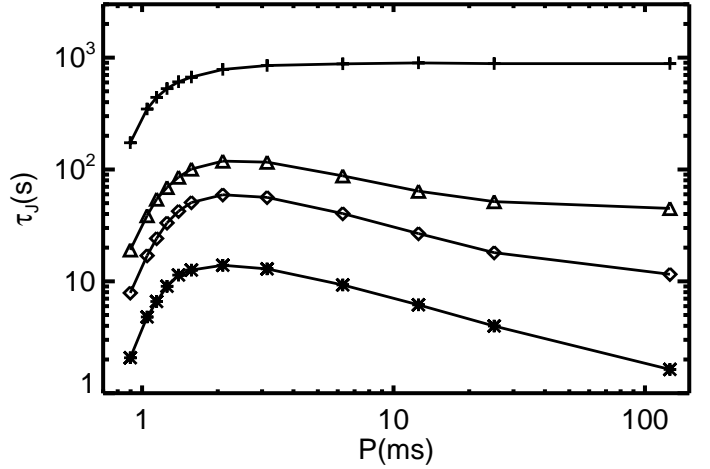,width=9.5cm}}}
\figcaption[x]{Spin-down timescale $\tau_{{\rm J}}\equiv \Omega/\dot{\Omega}$ as a function of rotation period $P$ for $L_{\bar{\nu}_{e}} = 8\times 10^{51}$ ergs s$^{-1}$ and 4 monopole surface magnetic field strengths: $B_{\nu} = 10^{13}$ G (cross), $10^{14}$ G (triangle), $2.5\times 10^{14}$ G (diamond), and $10^{15}$ G (asterisk).  The decrease in $\tau_{{\rm J}}$ for rapid rotation is due to the exponential enhancement in $\dot{M}$ for P $\lesssim$ 2-3 ms (eq.~[\ref{mdotcent}]).  The surface dipole field $B_{\nu}^{{\rm dip}}$ associated with the effective monopole field $B_{\nu}$ scales as $B_{\nu}^{{\rm dip}} \propto B_{\nu}P$ (see eq.~[\ref{bmoneff}]), which implies that the true dipole field appropriate to this figure can be much greater than $B_{\nu}$ for large $P$.  Note that an approximate analytic expression for $\tau_{{\rm J}}$ in the magnetically-driven limit ($P\lesssim P_{{\rm FMR}}$; eq.~[\ref{omegac}]) is given in equation (\ref{taujfmr}).
\label{plot:tauj}}
\end{figure}

The rate at which angular momentum is extracted from the PNS is $\dot{J} = I\dot{\Omega} = \Omega R_{A}^{2}\dot{M}$, where $J=I\Omega$ is the angular momentum of the PNS and $I \simeq (2/5)MR_{\nu}^2$ is the PNS moment of inertia.  Hence, given the Alfv\'{e}n radius from equation (\ref{ra}), the spin-down time of the PNS ($\tau_{J} \equiv \Omega/\dot{\Omega})$ in the non-relativistic, magnetically-driven limit is found to be 
\be 
\tau_{{\rm J}}|_{{\rm FMR}} \simeq 440\,\eta^{-1}\,B_{14}^{-4/3}\dot{M}_{-3}^{-1/3}P_{{\rm ms}}^{-2/3}\,{\rm s}.
\label{taujfmr}
\ee
This result shows explicitly that while increasing the mass-loading places a greater strain on the field lines ($R_{A} \propto \dot{M}^{-1/3}$) and hence reduces the net loss of angular momentum per gram $(\mathcal{L} = \Omega R_{A}^{2}\propto \dot{M}^{-2/3}$), the additional mass loss carries enough total angular momentum to increase the overall spin-down rate.  The high mass loss accompanying the Kelvin-Helmholtz epoch can therefore efficiently extract the rotational energy of the PNS.  

Figure \ref{plot:tauj} shows $\tau_{{\rm J}}$ calculated directly from our wind solutions as a function of $P$ at $L_{\bar{\nu}_{e},51} = 8$ for several magnetic field strengths (see also Table \ref{table:solutiontable}).  Our numerical results agree well with the analytic estimates in equation (\ref{taujfmr}) for winds that are in the FMR limit (i.e., $P \lesssim P_{{\rm FMR}}$; eq.~[\ref{omegac}]).  The sharp decline in $\tau_{{\rm J}}$ at short $P$ is due to the fact that $\tau_{{\rm J}} \propto \dot{M}^{-1/3}$ and that $\dot{M}$ is enhanced by centrifugal flinging for rapid rotation.  Note, however, that only for $B_{\nu} \gtrsim 10^{15}$ G is $\tau_{{\rm J}} \sim \tau_{{\rm KH}} \sim 10$ s for a millisecond rotator.   For $B_{\nu} = 10^{13}$ G and large $P$ the solutions are primarily thermally-driven and $\tau_{{\rm J}}$ is independent of $P$; this occurs because both the Alfv\'{e}n radius $R_{A}$ and mass-loss rate $\dot{M}$ (and hence $\tau_{{\rm J}}$) are independent of rotation rate for thermally-driven winds. 

\begin{figure}
\centerline{\hbox{\psfig{file=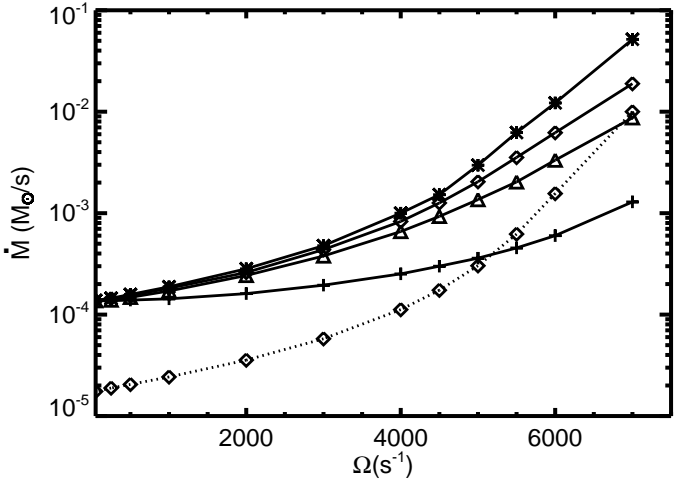,width=9.5cm}}}
\figcaption[x]{Mass loss rate $\dot{M}$ as a function of the rotation rate $\Omega$ at $L_{\bar{\nu}_{e}} = 8\times 10^{51}$ ergs s$^{-1}$ for $B_{\nu} = 10^{13}$ G (cross), $B_{\nu} = 10^{14}$ G (triangle), $2.5\times 10^{14}$ G (diamond), and $10^{15}$ G (asterisk); also shown is $\dot{M}(\Omega)$ for $L_{\bar{\nu}_{e}} = 3.5\times 10^{51}$ ergs s$^{-1}$ and $B_{\nu} = 2.5\times 10^{14}$ G (dotted, diamond).  $\dot{M}$ increases with increasing $\Omega$ and $B_{\nu}$ because centrifugal support expands the hydrostatic atmosphere (see Fig.~\ref{plot:density}).  For sufficiently large $B_{\nu}$ ($\gtrsim B_{\rm{cf}}$; eq.~[\ref{bcrit}]), however, $\dot{M}(\Omega)$ no longer increases with increasing $B_{\nu}$ because the wind corotates past the sonic point.  An approximate fit to the numerical results in this limit is given by equation (\ref{mdotcent}).  \\
\label{plot:mdotomega}}
\end{figure}

The Alfv\'{e}n radius and spin-down times calculated in equations (\ref{ra}) and (\ref{taujfmr}) depend on the mass loss rate $\dot{M}$ from the PNS, which itself depends on the PNS's rotation rate and magnetic field strength.  Figure \ref{plot:mdotomega} shows our determination of $\dot{M}$ as a function of $\Omega$ for field strengths $B_{\nu} = 10^{13},\, 10^{14},\, 2.5\times 10^{14},$ and $10^{15}$ G at $L_{\bar{\nu}_{e},51} = 8$, and $B_{\nu} = 2.5\times 10^{14}$ G at $L_{\bar{\nu}_{e},51} = 3.5$.  The mass loss rate increases rapidly with rotation for $P\lesssim$ 2-3 ms and also increases with $B_{\nu}$, though it saturates for the largest magnetic field strengths, as can be seen by comparing the $B_{\nu} = 2.5\times 10^{14}$ G and $10^{15}$ G solutions.  For sufficiently large $B_{\nu}$, such that $R_{A} \gtrsim R_{{\rm s}}$, we find empirically that $\dot{M}$ is given by
\be 
\dot{M} \simeq \dot{M}_{{\rm NRNM}}\exp{[(\Omega/\Omega_{{\rm cf}})^{2}]}\equiv \dot{M}_{{\rm cf}}, 
\label{mdotcent}
\ee
where $\dot{M}_{{\rm NRNM}}$ is the mass-loss rate for NRNM winds (eq.~[$\ref{mdotnrnm}$]) and $\Omega_{{\rm cf}} \approx 2700(L_{\bar{\nu}_{e},51}/8)^{0.08}$ s$^{-1}$.

\begin{figure}
\centerline{\hbox{\psfig{file=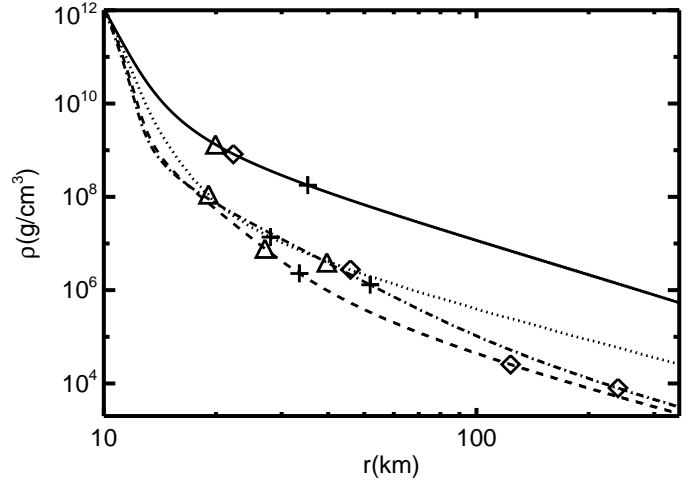,width=9.5cm}}}
\figcaption[x]{The density profiles in the inner $\sim$ 100 km for magnetically-driven wind solutions at $L_{\bar{\nu}_{e},51} = 8$, $B_{\nu} = 10^{15}$ G, $\Omega$ =  7000 s$^{-1}$ (solid), 5000 s$^{-1}$ (dotted), 2000 s$^{-1}$ (dashed), and 500 s$^{-1}$ (dot-dashed).  Triangles, crosses, and diamonds mark the slow point, the radius where $T=0.5$ MeV, and the Alf\'{e}n radius, respectively.  For all of the solutions, the supersonic portion of the density profile is altered from that of a thermally-driven solution due to magnetic acceleration.  For solutions with $\Omega \gtrsim 3000$ s$^{-1}$ the density scale height in the subsonic portion of the wind (interior to the triangle) is significantly larger due to centrifugal support.  This enhances the mass loss rate as shown in Figure \ref{plot:mdotomega}.
\label{plot:density}}
\end{figure}

\begin{figure}
\centerline{\hbox{\psfig{file=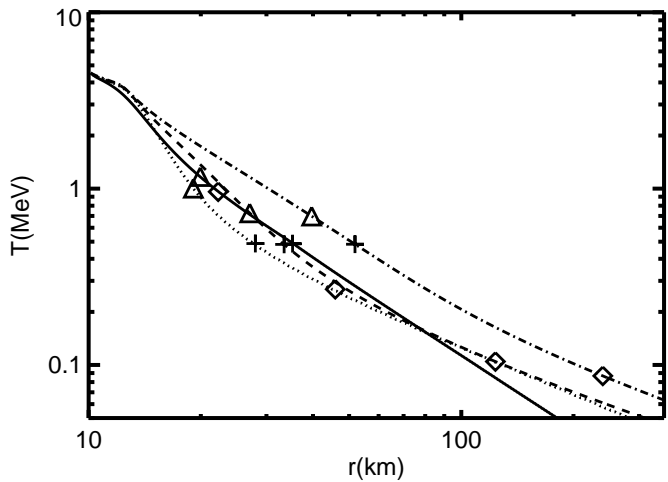,width=9.5cm}}}
\figcaption[x]{The temperature profiles in the inner $\sim$ 100 km for magnetically-driven wind solutions at $L_{\bar{\nu}_{e},51} = 8$, $B_{\nu} = 10^{15}$ G, $\Omega$ =  7000 s$^{-1}$ (solid), 5000 s$^{-1}$ (dotted), 2000 s$^{-1}$ (dashed), and 500 s$^{-1}$ (dot-dashed).  Triangles, crosses, and diamonds mark the slow point, the radius where $T=0.5$ MeV, and the Alf\'{e}n radius, respectively.  Notice that in all but the most rapidly-rotating case the $T=0.5$ MeV radius is located between the slow point and the Alfv\'{e}n radius, which implies that there is significant magnetic acceleration of the wind at the $T=0.5$ MeV radius.  As shown in Figure \ref{plot:tdynplot}, this significantly reduces the dynamical time at $T=0.5$ MeV, making magnetically-driven PNS winds more favorable for $r$-process nucleosynthesis than thermally-driven winds (see $\S$\ref{section:rprocess}).
\label{plot:temperature}}
\end{figure}

The enhanced mass loss shown in Figure \ref{plot:mdotomega} is due to the effect of strong magnetic fields and rapid rotation on the subsonic, hydrostatic structure of PNS winds.  Figures \ref{plot:density} and \ref{plot:temperature} show the density and temperature profiles, respectively, for winds with $B_{\nu} = 10^{15}$ G and $L_{\bar{\nu}_{e},51} = 8$ at several rotation rates.  For the most rapidly rotating solution (solid line), Figure \ref{plot:density} shows that centrifugal support is sufficient to expand the scale height of the hydrostatic atmosphere at small radii, resulting in the much higher mass loss rates seen in Figure \ref{plot:mdotomega}.  Analytically, we expect the centrifugal support to be important when $R_{\nu}\Omega \gtrsim c_{{\rm s},\nu}$, where $c_{{\rm s},\nu}$ is the sound speed at the PNS neutrinosphere.  We find that the inner sound speed depends only weakly on the neutrino luminosity: $c_{{\rm s},\nu} \approx 0.12(L_{\bar{\nu}_{e},51}/8)^{0.08}$ c, and thus that mass loss is enhanced for
\be 
\Omega \gtrsim 3600(L_{\bar{\nu}_{e},51}/8)^{0.08}\,{\rm s}^{-1}.
\label{omegacf}
\ee  
This region is denoted ``$\dot{M}$ Enhanced'' in Figure \ref{plot:regimes}.   Equation (\ref{omegacf}) is in good agreement with our numerically determined value of $\Omega_{{\rm cf}}$ defined in equation (\ref{mdotcent}).  

The enhancement of $\dot{M}$ implied by equations (\ref{mdotcent}) and (\ref{omegacf}) does not explicitly depend on the magnetic field strength.  However, centrifugal support of the PNS atmosphere only occurs if the field can sustain corotation out to the sonic radius $R_{{\rm s}}$; otherwise rotation has much less of an effect on the mass loss rate (see the solutions with low $B_{\nu}$ in Fig.~\ref{plot:mdotomega}).  The requirement of corotation out to $R_{{\rm s}}$ can be written as $R_{A}>R_{{\rm s},{\rm cf}}$ (eq.~[\ref{rcf}]), which in turn implies $B_{\nu} > B_{{\rm cf}}$, where
\be
  B_{{\rm cf}} \simeq 2\times 10^{14}\,\eta^{-3/4}\dot{M}_{-3}^{1/2}P_{{\rm ms}}^{1/2}\,{\rm G}. 
\label{bcrit}
\ee  
For $B_{\nu} \gtrsim B_{{\rm cf}}$, the mass loss from the PNS no longer increases with increasing $B_{\nu}$ (see Fig. \ref{plot:mdotomega}) because the wind already co-rotates out to the sonic radius where $\dot{M}$ is set.  In deriving $B_{{\rm cf}}$ in equation (\ref{bcrit}), we have used equation (\ref{ra}) for $R_{A}$ because, under most conditions, a wind that is centrifugally-supported will automatically be in the FMR limit, although the converse is not necessarily true.  We note that even relatively weakly-magnetized PNS's that are rapidly rotating will experience some degree of enhanced mass loss.  For instance, for $\Omega = 7000$ s$^{-1}$, even the lowest field strength solution in Figure \ref{plot:mdotomega} ($B_{\nu} = 10^{13}$ G) has a mass-loss rate almost an order of magnitude larger than its NRNM value.

The non-relativistic calculations we have presented here are only applicable for magnetizations $\sigma < 1$; this is equivalent to requiring $R_{A} < R_{{\rm L}} \equiv c/\Omega$, the radius of the light cylinder.  For $\sigma > 1$ the PNS wind becomes relativistic and its spin-down properties will change (see $\S$\ref{section:evolution} for a discussion).  Using equation (\ref{mdotcent}) for $\dot{M}$ (i.e., assuming $B_{\nu} > B_{{\rm cf}}$), the magnetization is given by
\be 
\sigma \simeq 0.05 B_{14}^{2}P_{{\rm ms}}^{-2}(L_{\bar{\nu}_{e},51}/8)^{-2.5}\exp[-5.4P_{{\rm ms}}^{-2}(L_{\bar{\nu}_{e},51}/8)^{-0.16}]. 
\label{sigma}
\ee
The $\sigma = 1$ boundary is denoted by a dotted line in Figure \ref{plot:regimes}.  PNSs with $\sigma > 1$ and $L_{\bar{\nu}_{e},51} \gtrsim 0.1$ will experience a relativistic phase accompanied by significant mass loss; this mass loss keeps the wind mildly relativistic, in contrast to the much higher $\sigma$ spin-down that will commence following $\tau_{{\rm KH}}$ ($L_{\bar{\nu}_{e},51} \ll 0.1$).  

\section{Applications and Discussion}
\label{section:discussion}

\subsection{Magnetized PNS Evolution}
\label{section:evolution}

With our numerical results in hand that sample a wide range of PNS wind conditions, we can begin to address the time evolution of a cooling, magnetized PNS.  In the early stages following the launch of SN shock the PNS is likely hot and inflated, with a radius exceeding the value of $R_{\nu} = 10$ km that we have assumed in all of the calculations presented in this paper.  This early phase is likely to be thermally-driven for all but the most highly-magnetized proto-magnetars, and, if a dynamo is at work, the large-scale field itself may still be amplifying during this phase (Thompson $\&$ Duncan 1993).  Using the collapse calculations of Bruenn, De Nisco, $\&$ Mezzacappa (2001) (from a $15$ M$_{\sun}$ progenitor of Woosley $\&$ Weaver 1995) T01 fit an approximate functional form to the PNS radial contraction:  $R_{\nu}\propto t^{-1/3}$ such that $R_{\nu}($1 s$) \simeq 15$ km and $R_{\nu}($2 s$) \simeq 12$ km.  The recent SN simulations of Buras et al.~(2003, 2006) with Boltzmann neutrino transport find a similar neutrinosphere radius at $t \sim$ 1\,s after bounce.

We do not attempt to address the uncertainties in early-time PNS cooling calculations, especially in the presence of large fields and rapid rotation.  Rather, we assume that the PNS has cooled to its final radius and completely established its global field by a time $t_{0} \sim 1$ s following core-collapse; we can then use the calculations presented in this paper to investigate the subsequent evolution of the PNS.  Following $t_{0}$ we assume a simplified PNS cooling evolution (similar to that used in TCQ, motivated by Figure 14 of Pons et al.~1999):
\begin{eqnarray} L_{\bar{\nu}_{e},51}(t) = L_{0}\left(\frac{t}{\tau_{{\rm KH}}}\right)^{-\delta} : t_{0}  < t < \tau_{{\rm KH}} \nonumber \\
L_{\bar{\nu}_{e},51}(t) = L_{0}\,\exp[-(t-\tau_{{\rm KH}})/\tau_{{\rm KH}}] : t > \tau_{{\rm KH}},
\label{lumevo}
\end{eqnarray}
where, for definitiveness in what follows, we take $L_{0}= 0.2$, $t_{0} = 1$ s, $\tau_{{\rm KH}} = 40$ s, and $\delta = 1$.  This cooling evolution is approximate because magnetar-strength fields and rapid rotation could alter $\tau_{{\rm KH}}$ or the form of the cooling profile (e.g., $\delta$) by affecting the neutrino opacity or the dynamics of the contraction itself (e.g., Villain et al.~2004; Dessart et al.~2006).  For instance, in 1D collapse calculations with rotation, Thompson, Quataert, $\&$ Burrows (2005) found that for $P\sim$ 1 ms, the total neutrino luminosity at $t \simeq 0.6$ s after bounce is $\sim 50\%$ smaller than in a non-rotating PNS.  

The dominant uncertainty in applying our results to magnetized PNS evolution is that we have assumed a monopole field geometry.  To relate our results to more realistic dipole simulations, we use the recent axisymmetric, relativistic MHD simulations of B06, who simulate neutron star spin-down for $\sigma \approx 0.3-20$. B06 show that the energy and angular momentum loss rates from aligned dipole spin-down can be described accurately by monopole formulae provided they are normalized to just the open magnetic flux; for instance, we can accurately apply our results for $\tau_{{\rm J}}$ (eq.~ [\ref{taujfmr}]) with a suitable renormalization of $B_{\nu}$.  

To apply the results of B06 we need to estimate the open magnetic flux in PNS winds.  In force-free spin-down calculations motivated by pulsars it is generally assumed that the radius of the last closed magnetic field line (the ``Y point'' $R_{Y}$) is coincident with the light cylinder (Contopoulos et al.~1999; Gruzinov 2005) so that the ratio between the fraction of open magnetic flux in the dipole and monopole cases is $R_{\nu}/R_{Y}=R_{\nu}/R_{{\rm L}}$.  This assumption is supported by force-free simulations (Spitkovsky 2006; McKinney 2006), which show that when mass-loading is completely negligible ($\sigma\rightarrow \infty$), $R_{Y} \simeq R_{{\rm L}}$.  However, B06 show that rapidly rotating, mildly mass-loaded MHD winds have a larger percentage of open magnetic flux than vacuum or force-free spin-down would imply (i.e., $R_{Y} < R_{{\rm L}}$).  From Table 4 of B06 we fit the approximate power law 
\be 
R_{Y}/R_{{\rm L}} \simeq 0.31\sigma^{0.15}
\label{ryoverrl}
 \ee
for the range $\sigma \in [0.298,17.5]$ and for a fixed rotation period of order one millisecond.  Although there are uncertainties in quantitatively extrapolating B06's results, reaching the pure force-free limit with $R_{Y} \simeq R_{{\rm L}}$ appears to require $\sigma \gg 1$.  We therefore conclude that magnetized, rapidly rotating PNS winds (with $\sigma \in \{10^{-3},10^{3}\}$ for $t<\tau_{{\rm KH}}$ under most circumstances) will typically possess excess open magnetic flux. 

Because the results of B06 for $R_{Y}/R_{{\rm L}}$ cover only a relatively narrow portion of PNS parameter-space we must proceed with caution in generalizing their results to our calculations; on the other hand, their basic conclusion shows a weak dependence on $\sigma$ and $\Omega$, and has a solid theoretical explanation (Mestel $\&$ Spruit 1987).  Hence, we have attempted to apply the results of B06 to gain insight into the multi-dimensional generalization of our calculations, but we check at every step that our calculations are not overly sensitive to extrapolations of B06's results.  Combining equation (\ref{ryoverrl}) with the empirical formula B06 provide (their eq.~[25]) relating the effective monopole surface field $B_{\nu}^{{\rm mon}}$ to the true equatorial dipole field $B_{\nu}^{{\rm dip}}$ we find
\be 
B_{\nu}^{{\rm mon}} \simeq 0.6B_{\nu}^{{\rm dip}}\frac{R_{\nu}}{R_{{\rm L}}}\left(\frac{R_{Y}}{R_{{\rm L}}}\right)^{-1} \simeq 0.4B_{\nu}^{{\rm dip}} P_{{\rm ms}}^{-1}\sigma^{-0.15},
\label{bmoneff}
\ee
which remains approximately valid for a substantial range in $\sigma$, provided we enforce $R_{{\rm Y}} = R_{{\rm L}}$ for $\sigma \gtrsim 10^{3}$ and keep $R_{Y} > R_{\nu}$.  

Using equation (\ref{bmoneff}) we substitute $B_{\nu}^{{\rm mon}}$ for $B_{\nu}$ in the results of $\S$\ref{section:results} and integrate from $t_{0}$ to $\tau_{{\rm KH}}$ using our fiducial cooling evolution (eq.~[\ref{lumevo}]) to obtain the total energy and mass loss during the Kelvin-Helmholtz epoch ($E_{{\rm tot}}^{a}$ and $\Delta M_{{\rm tot}}$, respectively) as a function of the fixed dipole surface field $B_{\nu}^{{\rm dip}}$ and for initial rotation periods $P_{0} \in \{1,10\}$ ms.  Although a neutron star will eventually impart any remaining rotational energy after $\tau_{{\rm KH}}$ to its surroundings through an ultra-relativistic, pulsar-like outflow, we concentrate on the wind evolution prior to the end of the Kelvin-Helmholtz phase because our calculations are primarily suited to studying mass-loaded spin-down and because we are interested in energy that can be extracted sufficiently early to affect the rapidly outward-propagating SN shock.  In performing our calculations when the wind is relativistic ($\sigma > 1$), we continue using equation (\ref{mdotcent}) for $\dot{M}$ and employ the relativistic spin-down formula given by B06 (their eq.~[26]):
\be 
\dot{E}^{a}_{{\rm REL}} \simeq 1.5\times 10^{47}(B_{14}^{{\rm dip}})^{2}P_{{\rm ms}}^{-4}\left(\frac{R_{Y}}{R_{{\rm L}}}\right)^{-2}\,{\rm ergs}\,\,{\rm s}^{-1}, 
\label{edotrel}
\ee
evaluated for $R_{Y}/R_{{\rm L}}$ given by equation (\ref{ryoverrl}).  

Although it is possible that equation (\ref{ryoverrl}) may not be accurate far outside the parameter regime B06 considered, we found that re-running the calculations at fixed $R_{{\rm Y}}/R_{{\rm L}} = 1/3$ changed $E^{a}_{{\rm tot}}$ by, at most, a factor of a few.  A similar uncertainty in our calculations is that we have used our equatorial $\dot{M}$ over all $4\pi$ steradian; thus, we have probably overestimated $\Delta M_{{\rm tot}}$ by a factor of $\sim$ 2 due to the effects of closed magnetic flux and because centrifugal flinging concentrates $\dot{M}$ at low latitudes.  Although a direct comparison of the mass loss rate between our solutions and the dipole simulations of B06 is difficult, we find that the dependence of $\dot{M}$ on $\Omega$ is similar between our solutions when the surface temperature in B06's simulations is scaled to an appropriate neutrino luminosity.

\begin{figure*}
\centerline{\hbox{\psfig{file=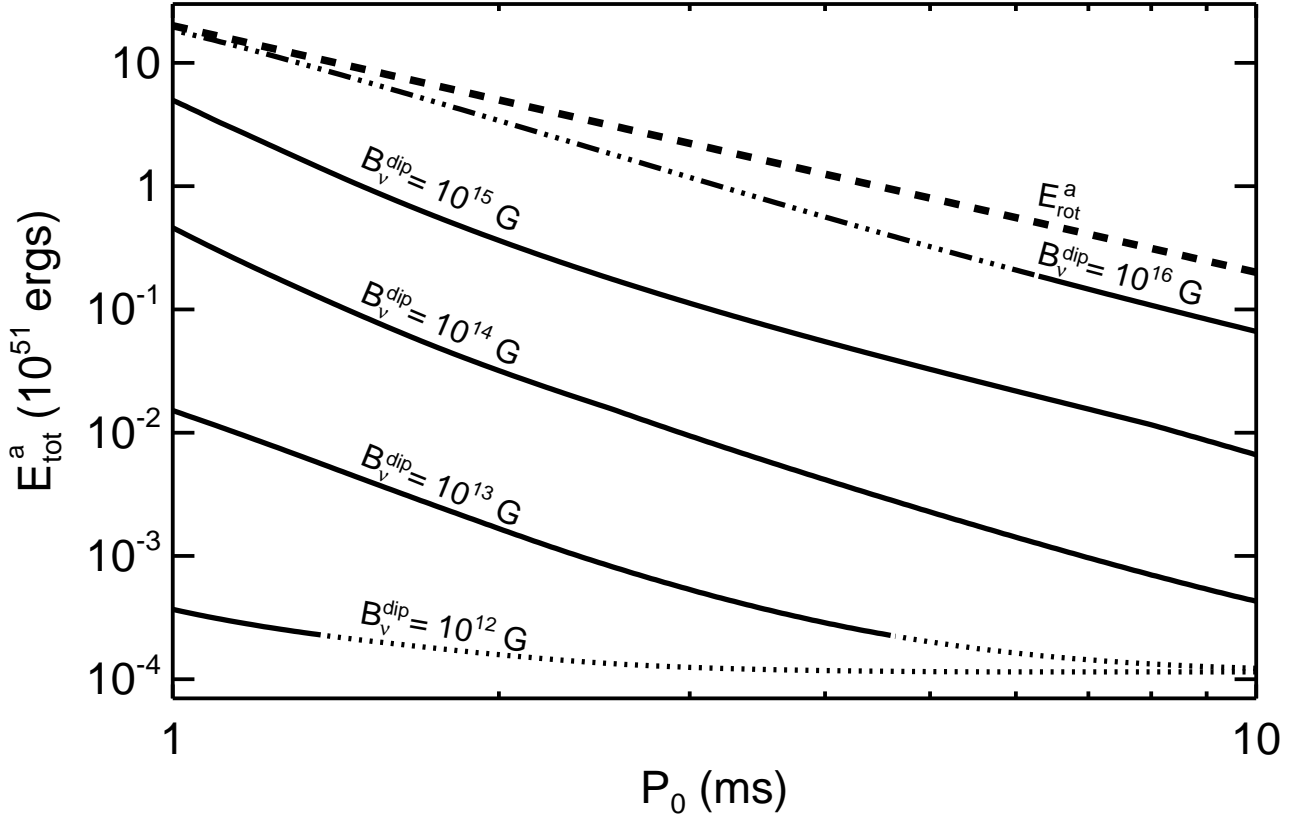,width=18cm}}}
\figcaption[x]{$E^{a}_{{\rm tot}}$ as a function of the PNS initial period $P_{0}$ for fixed dipole surface fields that range from those typical of rotation-powered pulsars to those capable of producing hyper-energetic SNe.  $E^{a}_{{\rm tot}}$ is the total energy carried by the PNS wind at infinity, calculated by evolving the PNS from time $t_{0} = 1$ s to $\tau_{{\rm KH}}=40$ s, assuming the PNS cooling evolution of equation (\ref{lumevo}) and using results from the aligned dipole simulations of B06 to relate $B_{\nu}^{{\rm dip}}$ to our monopole calculations (see eq.~[\ref{bmoneff}]).   The line style denotes the wind phase that dominates the total energy loss (dotted = thermally-driven; solid = non-relativistic, magnetically-driven; 3 dot-dash = relativistic, magnetically-driven).  The dashed line at the top shows the total initial rotational energy of the PNS.  For $B_{\nu}^{{\rm dip}} = 10^{12}$ G neutrino-heated, thermally-driven outflow dominates for $P_{0}\gtrsim 1$ ms, while for $B_{\nu}^{{\rm dip}} = 10^{16}$ G almost the entire rotational energy of the PNS is extracted by a relativistic, magnetized wind during the Kelvin-Helmholtz phase.
\label{plot:etot}}
\end{figure*}

\begin{figure}
\centerline{\hbox{\psfig{file=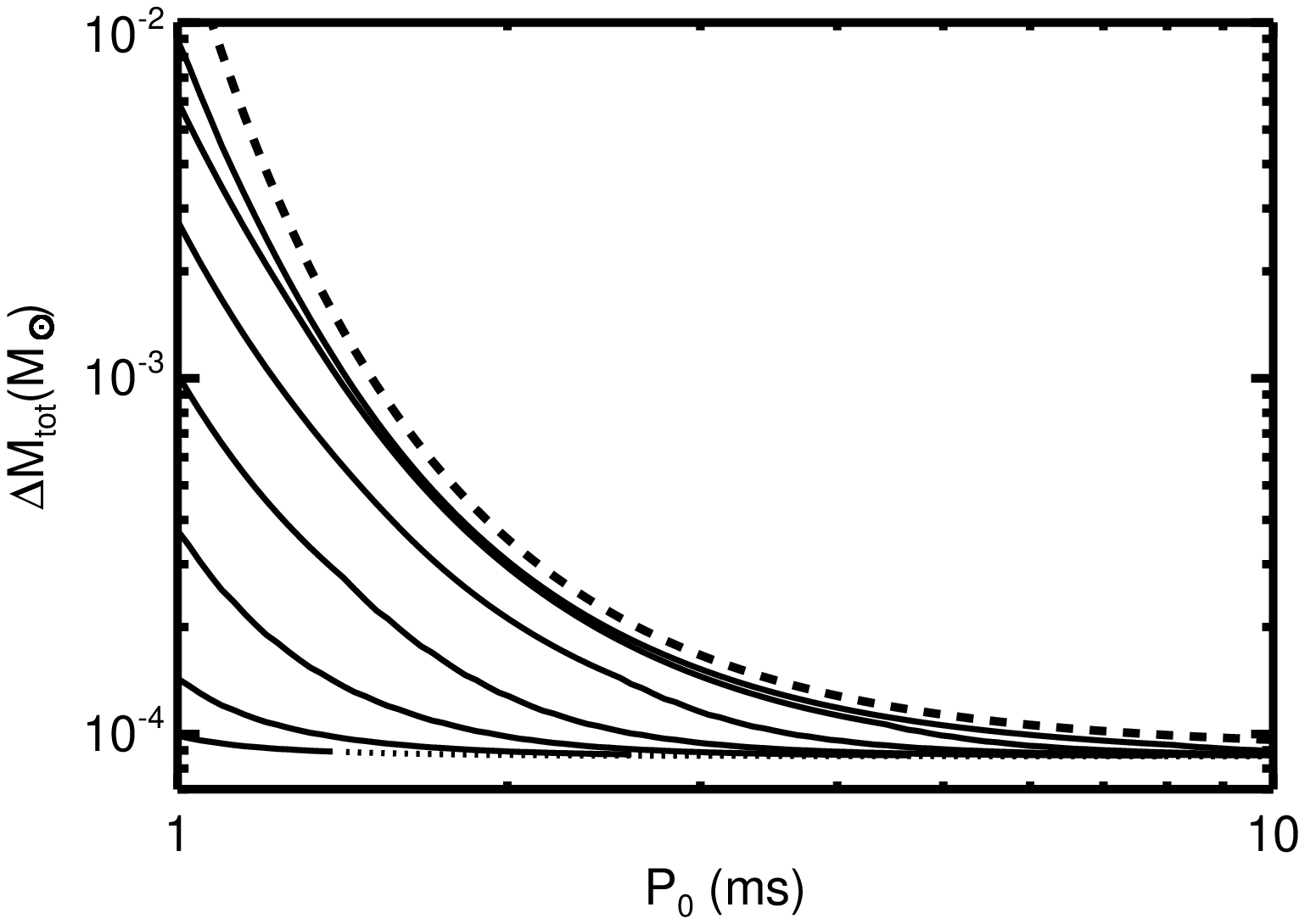,width=9.5cm}}}
\figcaption[x]{The total mass extracted via a PNS wind ($\Delta M_{{\rm tot}}$) from time $t_{0} = 1$ s to $\tau_{{\rm KH}}=40$ s as a function of the PNS initial period $P_{0}$ for dipole surface fields $B_{\nu}^{{\rm dip}} = 10^{12}, 3\times 10^{12}, 10^{13},3\times 10^{13}, 10^{14}, 3\times 10^{14},$ and $10^{15}$ G (from bottom to top).  These calculations assume the PNS cooling evolution of equation (\ref{lumevo}) and use the aligned dipole simulations of B06 to relate $B_{\nu}^{{\rm dip}}$ to our monopole calculations (see eq.~[\ref{bmoneff}]).  The line style denotes the wind phase that dominates the total energy loss (dotted = thermally-driven; solid = non-relativistic, magnetically-driven; 3 dot-dash = relativistic, magnetically-driven).  The results of this figure demonstrate that for $P_{0}\lesssim$ 2$-$3 ms and $B_{\nu}^{{\rm dip}}\gtrsim 10^{13}$ G, centrifugal flinging enhances the total mass extracted from a PNS wind during $\tau_{{\rm KH}}$.  An analytic approximation to $\Delta M_{{\rm tot}}$ for high $B_{\nu}$ is given in equation (\ref{deltamtot}) and is shown with a thick dashed line.  For $P\lesssim3$\,ms, $\Delta M_{\rm tot}$ is essentially the same for the $B_\nu^{\rm dip}=3\times10^{14}$\,G and $10^{15}$\,G models because of the saturation of $\dot{M}$ at fixed $\Omega$ for $B \gtrsim B_{\rm cf}$.
\label{plot:mtot}}
\end{figure}

Our estimates for $E_{{\rm tot}}^{a}$ and $\Delta M_{{\rm tot}}$ during the Kelvin-Helmholtz phase are presented in Figures \ref{plot:etot} and \ref{plot:mtot} for values of $B_{\nu}^{{\rm dip}}$ ranging from $10^{12}-10^{16}$ G.  The single dominant phase contributing the majority of the energy extracted for a given initial rotation rate $P_{0}$ is denoted by the line style (thermally-driven [NRNM] = dotted; non-relativistic, magnetically-driven [FMR] = solid; relativistic, magnetically-driven [REL] = 3 dot-dash).  While all PNSs pass through each wind phase sometime during $\tau_{{\rm KH}}$ (see Fig.~\ref{plot:regimes}), PNSs can still be usefully classified into 3 types based on which wind phase dominates the total energy loss during $\tau_{{\rm KH}}$:

\noindent(1) ${\bf Thermally-Driven\,Winds}\,\,{\bf (B_{14}^{{\rm dip}} \lesssim 10^{-2}P_{{\rm ms}}^{2}):}$ For low magnetic field strengths, Figures \ref{plot:etot} and \ref{plot:mtot} show that the total mass and energy loss are effectively at the NRNM values for a large range of initial periods:  $\Delta M_{{\rm tot}}^{{\rm NRNM}} \simeq 10^{-4}(t_{0}/1{\rm s})^{-1.5} $M$_{\sun}$, $E_{{\rm tot}}^{{\rm NRNM}} \simeq 2\times 10^{47}(t_{0}/1{\rm s})^{-2.2}$ ergs.  Energy and mass loss from this class of PNSs is generally modest and is dominated by early times.  Analysis of the Parkes multibeam survey suggests that half of all pulsars are born with $B_{\nu}^{{\rm dip}} < 3\times 10^{12}$ G (Vranesevic et al.~2004), implying that NRNM winds dominate the majority of neutron star births, independent of the birth-period distribution.  Spin-down during $\tau_{{\rm KH}}$ is negligible for PNSs of this type and the supernova remnants associated with the production of NRNM PNSs will not be significantly modified by the small energy injected during the cooling phase.

\noindent(2) ${\bf Non-Relativistic,\,Magnetically-Driven\,Winds} \\ {\bf (10^{-2}P_{{\rm ms}}^{2} \lesssim B_{14}^{{\rm dip}} \lesssim 2P_{{\rm ms}}^{2}\exp[2P_{{\rm ms}}^{-2}]):}$  For $\sim 10^{13}-10^{15}$ G surface field strengths, Figure \ref{plot:etot} shows that $E_{{\rm tot}}^{a}$ is dominated by a non-relativistic, magnetically-driven outflow during $\tau_{{\rm KH}}$ for most periods between 1 and 10 ms.  Note that most observed Galactic magnetars have field strengths in this range (Kouveliotou et al.~1998).  Figure \ref{plot:etot} shows that for $B \sim 10^{15}$ G and $P \lesssim$ 2 ms, more than $10^{51}$ ergs can be lost to a non-relativistic, magnetically-driven outflow during $\tau_{KH}$, and that, over a broad range of initial spin period, the energy extracted is many times larger than from a slowly rotating PNS.  Because non-relativistic outflows are efficiently collimated along the rotational axis by magnetic stresses (B06), the energy per unit solid angle at the pole may exceed that of the SNe, potentially altering the SNe shock's morphology and nucleosynthetic yield. 

If, for the purposes of an analytic estimate, we assume that $R_{Y} = R_{{\rm L}}/3$ then the total energy extracted for this class of PNS can be approximated as\footnote{Equation (\ref{etotfmr}) also assumes that $\delta = 1$, that strict corotation can be maintained by the magnetic field (this criteria is well-satisfied because late times dominate the energy release, and winds at late times have lower $\dot{M}$ and are easier to support magneto-centrifugally), and that $\Omega$ does not evolve significantly (i.e., $\tau_{J} \gtrsim \tau_{{\rm KH}}$); this is well-satisfied for field-strengths at which the outflow energy is indeed extracted via a non-relativistic outflow.}
\begin{eqnarray}
E_{{\rm tot}}^{{\rm FMR}}\simeq 10^{50}(B_{14}^{{\rm dip}})^{4/3}P_{{\rm ms}}^{-8/3} \nonumber \\
\times[\tau_{{\rm f}}^{1/6}-t_{0}^{1/6}]\exp[1.3 P_{{\rm ms}}^{-2}]\,{\rm ergs}, 
\label{etotfmr}
\end{eqnarray}
where $\tau_{{\rm f}} \equiv {\rm min}$$\{\tau_{{\rm KH}},\tau_{{\rm REL}}\}$, $t_{0}$ and $\tau_{{\rm f}}$ are in seconds, and $\tau_{{\rm REL}}$ $\simeq 7$ s $(B_{14}^{{\rm dip}})^{-0.8}P_{{\rm ms}}^{1.6}\exp[2.2P_{{\rm ms}}^{-2}]$ is the time after which the wind becomes relativistic.  The weak dependence of equation (\ref{etotfmr}) on $\tau_{{\rm KH}}$ and $t_{0}$ shows that the total energy extracted in the magnetically-dominated phase is relatively insensitive to our choice for the PNS thermal evolution because the energy loss is distributed almost equally per decade in time.  For the same reason, $E_{{\rm tot}}^{{\rm FMR}}$ is relatively insensitive to the precise time during the Kelvin-Helmholtz epoch at which the PNS cools to its final radius.  

Figure \ref{plot:mtot} shows that the total mass loss from PNSs is enhanced for $P\lesssim 3$ ms and $B_{\nu}^{{\rm dip}} \gtrsim 10^{13}$ G; the total mass loss increases with $B_{\nu}^{{\rm dip}}$, saturating for $B_{\nu}^{{\rm dip}} \gtrsim 3\times 10^{14}$ G (i.e., $B_{\nu} \gtrsim B_{{\rm cf}}$; see eq.~[\ref{bcrit}]).  Assuming no evolution of $\Omega$, the integrated mass loss is approximately
\be 
\Delta M_{{\rm tot}} \simeq 10^{-4}\,t_{0}^{-1.5}\exp[5.4 P_{{\rm ms}}^{-2}]\,{\rm M}_{\sun} 
\label{deltamtot}
\ee
for large $B_{\nu}$.  For $P_{0}\approx 1$ ms and $P_{0}\approx 3$ ms, the total mass extracted is thus $\sim 10^{2}$ and $\sim$ 2 times greater, respectively, than from a non-rotating, non-magnetic PNS.  Equation (\ref{deltamtot}) is shown with a dashed line in Figure \ref{plot:mtot}.

\noindent(3) ${\bf Relativistic,\,Magnetically-Driven\,Winds\,} \\ {\bf (B_{14}^{{\rm dip}} \gtrsim 2P_{{\rm ms}}^{2}\exp[2P_{{\rm ms}}^{-2}]):}$ For $B_{\nu}^{{\rm dip}} = 10^{16}$ G and $P_{0} \lesssim 6$ ms, the energy loss from PNSs is dominated by a relativistic, magnetically-driven outflow during the Kelvin-Helmholtz phase.  As Figure \ref{plot:etot} illustrates, PNSs of this type lose $\approx 10-100\%$ of their total rotational energy during $\tau_{{\rm KH}}$.  For sufficiently rapid rotation such PNSs are therefore candidates for the central engine of hyper-energetic SNe and LGRBs (see $\S$\ref{section:LGRB}).  PNS winds of this type differ from the ultra-relativistic ($\sigma \gg 10^{3}$), pulsar-like phase that begins once neutrino heating completely subsides in two important ways: (1) neutrino heating causes significant mass loss that maintains modest $\sigma$ ($\lesssim 10^{2}-10^{3}$); and (2) mass-loss leads to more open magnetic flux (eq.~[\ref{ryoverrl}]) than in the pure force-free case and thus the PNS spins down more rapidly (see eq.~[\ref{edotrel}]).  

\subsection{Hyper-Energetic SNe and Long Duration Gamma-Ray Bursts}
\label{section:LGRB}

One observational manifestation of early energy loss from rapidly rotating proto-magnetars may be ``hyper-energetic'' SNe, which we define as having energy greater than a SN's usual $\sim 10^{51}$ ergs.  SNe significantly energized by proto-magnetar winds are naturally asymmetric given the preferred direction associated with the rotation axis of the PNS, and if energized sufficiently early, their nucleosynthetic yield may be appreciably modified; additionally, in some cases proto-magnetar winds may provide conditions favorable for long-duration GRBs within just seconds of the progenitor core collapse (Thompson 1994; Wheeler et al.~2000; TCQ).  In order to possess $\gtrsim 10^{51}$ ergs of rotational energy a PNS must be born with a period $P_{0} \lesssim 4$ ms, and, as Figure \ref{plot:etot} illustrates, such a large energy is extracted during the Kelvin-Helmholtz epoch only, for $B_{\nu}^{{\rm dip}} \gtrsim 10^{15}$ G, if the energetically-dominant form of outflow is at least mildly relativistic.  Using equation (\ref{edotrel}), the spin-down timescale for relativistic outflow from a proto-magnetar is
\be 
\tau_{{\rm J}}^{{\rm REL}} \simeq 350(B_{15}^{{\rm dip}})^{-2}P_{{\rm ms}}^{2}\left(\frac{R_{Y}/R_{{\rm L}}}{1/3}\right)^{2}{\rm s}. 
\label{tausrel}
\ee
When $R_{{\rm Y}} \approx  R_{{\rm L}}$ ($\sigma \gg 1$), equation (\ref{tausrel}) reduces to the canonical force-free (``vacuum dipole'') spin-down timescale, but as was discussed in $\S$\ref{section:evolution}, neutrino-heated mass flux (which is significantly enhanced for $P \sim$ 1 ms) maintains modest $\sigma$ at early times, and therefore the PNS may spin down up to an order of magnitude faster during the Kelvin-Helmholtz epoch (B06). 

For a surface dipole field typical of observed Galactic magnetars ($\sim 10^{14}-10^{15}$ G) the SN shock is only energized with $\gtrsim 10^{51}$ ergs during $\tau_{{\rm KH}}$ for initial periods $\lesssim 1-2$ ms; however, such a large total rotational energy ($\gtrsim 10^{52}$ ergs) probably cannot be typical of magnetar birth because even if it is not extracted during $\tau_{{\rm KH}}$ this rotational energy will eventually be transferred to the surrounding environment, and observations of Galactic magnetar SN remnants are inconsistent with such a large energy (e.g., Vink $\&$ Kuiper 2006)\footnote{One caveat to this argument is that a significant portion of the rotational energy could escape as gravitational waves in a time $\ll \tau_{{\rm J}}$ (see Stella et al.~2005).}.  In addition, the rate of hyper-energetic SNe is probably much smaller than the rate of magnetar births: Podsiadlowski et al.~(2004) estimate that the hyper-energetic SNe rate is only $\sim 0.01-0.1\%$ of the radio pulsar birthrate, while Woods \& Thompson (2004) estimate that the Galactic magnetar birthrate is at least $\sim 10\%$ of the radio pulsar birthrate.  

Consider, however, a rarer class of proto-magnetar with rapid initial rotation ($P_{0} \sim 1$ ms) and a somewhat stronger global magnetic field ($B_{\nu}^{{\rm dip}} \approx 3\times 10^{15}-10^{16}$ G; if a rapid birth period is indeed the cause of such a strong field, these assumptions are not independent).  Figure \ref{plot:etot} shows that winds from PNSs with these characteristics are dominated energetically by at least mildly relativistic outflow.  Evolving a proto-magnetar of this type with $P_{0}=1$ ms using the calculations described in $\S$\ref{section:evolution} we find that the total energy extracted during the Kelvin-Helmholtz phase is $1.1\times 10^{52}(1.8\times 10^{52}$) ergs for a surface dipole field strength of $3\times 10^{15}(10^{16}$) G; this represents $50\%(80\%)$ of the total rotational energy of the PNS.  We find that almost all of this energy is extracted with $\sigma < 10^{3}$.  More specifically, for $B_{\nu}^{{\rm dip}} = 3\times 10^{15}(10^{16})$ G, we find that $\approx 7\times 10^{51}(1.1\times 10^{52})$ ergs of rotational energy is extracted with $\sigma < 10$ in the first 5(2) s following the launch of the SN shock; the assumption of excess open magnetic flux over the pure force-free case (eq.~[\ref{ryoverrl}]) is therefore especially well-justified because wind solutions with precisely these parameters ($B^{{\rm dip}} \sim 10^{15}-10^{16}$ G, $P \sim$ 1 ms, and $\sigma \sim 0.1-10$) have been calculated by B06.  The rapid spin-down (eq.~[\ref{tausrel}]) and efficient energy extraction that occurs immediately following the birth of a magnetar of this type may energize the SN shock sufficiently rapidly to enhance its $^{56}$Ni nucleosynthetic yield (e.g., Nakamura et al.~2001), one of the observational signatures of hyper-energetic SNe (however, see Soderberg 2006).  We note, however, that the ability of proto-magnetar winds to affect the SN nucleosynthesis is sensitive to the evolution of the magnetic field and radius of the PNS, since the time for the PNS to contract to its final radius is similar to the time over which energy must be extracted to affect the $^{56}$Ni yield.

Although a significant portion of the PNS rotational energy emerges with $\sigma < 10$, which causes enhanced spin-down at early times, we find that the total energy extracted is distributed almost uniformly in log($\sigma$) in the range $\sigma \in \{0.1,1000\}$.  Indeed, for $B_{\nu}^{{\rm dip}} = 3\times 10^{15}(10^{16})$ G we find that $2\times 10^{51}(4\times 10^{51})$ ergs is extracted with $10 <\sigma < 100$ in the first 18(7) s, and that an additional $10^{51}(3\times 10^{51})$ ergs is extracted with $100 < \sigma < 1000$ by 39(32) s following the launch of the SN shock.  If the Poynting-Flux of this outflow can be efficiently converted to kinetic energy (e.g., Drenkhahn $\&$ Spruit 2002), then the high energy to mass density implied by the wind's large $\sigma$ will result in acceleration to a comparably large asymptotic Lorentz factor.  The fact that a significant portion of the PNS rotational energy emerges with $\sigma \sim 10-1000$ on a timescale $\tau_{{\rm J}}^{\rm {REL}} \sim \tau_{{\rm KH}}\sim 10-100$ s thus makes the birth of proto-magnetars with $P_{0} \sim 1$ ms and $B_{\nu}^{{\rm dip}} \sim 3\times 10^{15}-10^{16}$ G a viable candidate for the central engine of LGRBs.  Indeed, it is important to note that for initial PNS periods that give the right energetics for LGRBs ($P\approx 1-3$ ms) and surface magnetic fields that give the right timescale for LGRBs ($B_{\nu}^{{\rm dip}} \approx 3\times 10^{15}-10^{16}$ G) the magnetization of the resulting proto-magnetar wind - which is not a free parameter of the problem - is consistent with the Lorentz factors inferred from LGRBs ($\Gamma \sim 100$; Lithwick $\&$ Sari 2001; Granot \& Kumar 2005).  One shortcoming of our calculations, however, is that we cannot address whether proto-magnetar winds will have large-scale collimation, as suggested by observations of some LGRB afterglows (e.g., ``jet breaks''; Rhoads 1997, 1999; Frail et al.~2001).  As noted in the introduction, the collimation of the relativistic wind may depend on its interaction with less relativistic material (which is ejected earlier and is more likely to be collimated about the pole; B06) or the stellar mantle (e.g., Wheeler et al.~2000; Uzdensky $\&$ MacFadyen 2006).

\subsection{$r$-process Nucleosynthesis}
\label{section:rprocess}

PNS winds are a plausible candidate for the astrophysical source of heavy $r$-process nuclides (Woosley $\&$ Hoffman 1992; Meyer et al.~1992).  The chief requirement for the $r$-process in PNS winds successfully reaching the critical third abundance peak at $A \approx 195$ is that the ratio of neutrons to seed nuclei (the ``neutron-to-seed'' ratio) remain high until the outflow cools to the point at which $r$-process can commence ($T\approx$ 0.1 MeV).  As many previous investigations have emphasized (e.g., Hoffman, Woosley, $\&$ Qian 1997; hereafter HWQ), the primary wind properties necessary to achieve and maintain a large neutron-to-seed ratio are: (a) a high asymptotic wind entropy\footnote{More precisely the $r$-process requires a high wind entropy at $T=0.5$ MeV ($S_{0.5\,{\rm MeV}}$); however, entropy from neutrino heating saturates inside the radius where $T=0.5$ MeV so that $S_{0.5\,{\rm MeV}} \simeq S^{a}$(indeed, as discussed in $\S\ref{section:microphysics}$, we artificially set $\dot{q}_{\nu}=0$ for $T<0.5$ MeV).  In $\S\ref{section:waveheating}$ we will distinguish between $S_{0.5\,{\rm MeV}}$ and $S^{a}$ when we consider the more radially-extended effects that wave heating can have on the wind entropy.} $S^{a}$, (b) a small asymptotic electron fraction $Y_{e}^{a}$ (large neutron fraction), and (c) a short dynamical timescale $\tau_{{\rm dyn}}$, where we follow HWQ in defining $\tau_{{\rm dyn}}$ as\footnote{We caution that $\tau_{{\rm dyn}}$ is sometimes defined in terms of the density scale height and this dynamical timescale, under the radiation-dominated and approximately constant entropy conditions at $T\approx0.5$ MeV, is a factor of 3 shorter than that defined in equation (\ref{taudef}).}
\be 
\tau_{{\rm dyn}} \equiv \left[\frac{T}{v_{r}|dT/dr|}\right]_{T=0.5 {\rm MeV}} \sim \left[\frac{r}{v_{r}}\right]_{T=0.5 {\rm MeV}}, 
\label{taudef}
\ee
where the second equality only holds as an order-of-magnitude estimate.  The dynamical time is defined at $T=0.5$ MeV because this is the radius at which $\alpha$-particles, the building blocks of seed nuclei, first form.

\begin{figure}
\centerline{\hbox{\psfig{file=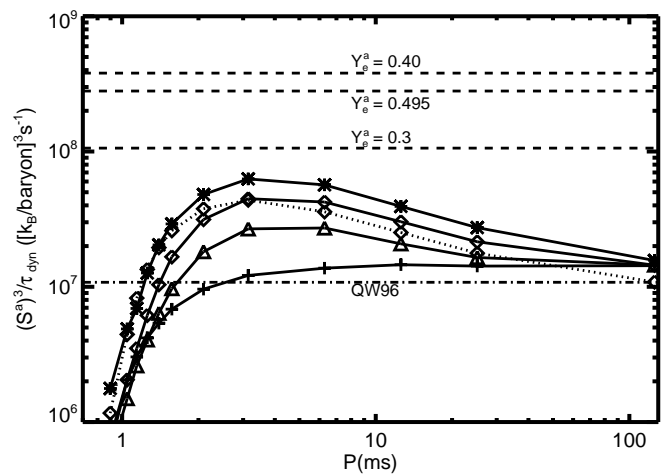,width=9.5cm}}}
\figcaption[x]{$(S^{a})^{3}/\tau_{{\rm dyn}}$ as a function of the rotation period $P$ for monopole magnetic field strengths $B_{\nu} = 10^{13}$ G (cross), $10^{14}$ G (triangle), $2.5\times 10^{14}$ G (diamond), and $10^{15}$ G (asterisk) at $L_{\bar{\nu}_{e}}=8 \times 10^{51}$ ergs s$^{-1}$, where $S^{a}$ is asymptotic wind entropy and $\tau_{{\rm dyn}}$ is the dynamical time evaluated at $T=0.5$ MeV (eq.~[\ref{taudef}]).  The ratio $(S^{a})^{3}/\tau_{{\rm dyn}}$ is also shown for $B_{\nu} = 2.5\times 10^{14}$ G at $L_{\bar{\nu}_{e}}=3.5 \times 10^{51}$ ergs s$^{-1}$ (diamond, dotted).  Shown with dashed lines are the approximate thresholds above which $r$-process can proceed to the third abundance peak ($A \approx 195$), taken from the numerical study of Hoffman, Woosley, $\&$ Qian 1997 (their Table 5), for several $Y_{e}^{a}$; notice that the threshold actually decreases with $Y_{e}^{a}$ for $Y_{e}^{a} \gtrsim 0.46$.   The dot-dashed line is $(S^{a})^{3}/\tau_{{\rm dyn}}$ calculated from the analytic expressions given by Qian $\&$ Woosley (1996) for NRNM winds at $L_{\bar{\nu}_{e}}=8 \times 10^{51}$ ergs s$^{-1}$ (their eqs.~[48a] and [61]).  This figure highlights that strongly magnetized, rapidly rotating PNS winds produce conditions significantly more favorable for successful third peak $r$-process; the optimal conditions obtain for $B_{\nu} \gtrsim 10^{14}$ G and $P \sim 2-10$ ms because $\tau_{{\rm dyn}}$ is reduced to a fraction of the rotation period by magneto-centrifugal acceleration (see Fig.~\ref{plot:tdynplot}).  The decrease in $(S^{a})^{3}/\tau_{{\rm dyn}}$ for $P\lesssim 3$ ms arises because, for sufficiently rapid rotation, magneto-centrifugal acceleration reduces the advection time of wind material through the heating region, thus decreasing the asymptotic entropy $S^{a}$(eq.~[\ref{sa}]).  \\
\label{plot:svsomega}}
\end{figure}

\begin{figure}
\centerline{\hbox{\psfig{file=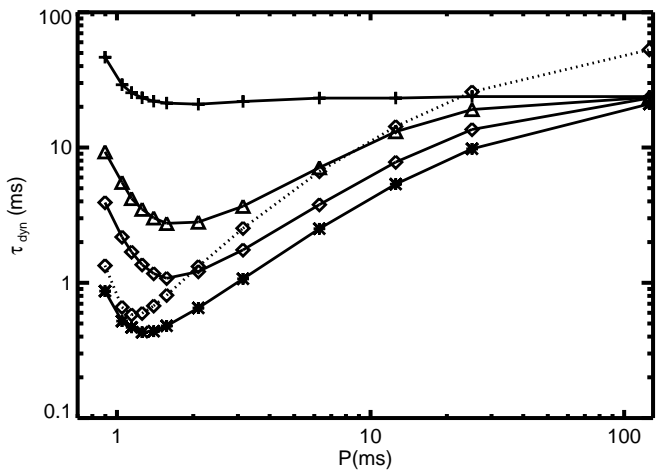,width=9.5cm}}}
\figcaption[x]{The dynamical time  $\tau_{{\rm dyn}}$ at the radius where $T=0.5$ MeV (eq.~[\ref{taudef}]) as a function of the PNS rotation period ($P$) at $L_{\bar{\nu}_{e}}=8 \times 10^{51}$ ergs s$^{-1}$ for monopole field strengths $B_{\nu} = 10^{13}$ G (cross), $10^{14}$ G (triangle), $2.5\times 10^{14}$ G (diamond), and $10^{15}$ G (asterisk).  The dynamical time is also shown for $B_{\nu} = 2.5\times 10^{14}$ G at $L_{\bar{\nu}_{e}}=3.5 \times 10^{51}$ ergs s$^{-1}$ (diamond, dotted).  For slow rotation and weak magnetic fields, $\tau_{{\rm dyn}}$ approaches a constant value ($\approx 20$ ms at $L_{\bar{\nu}_{e}}=8 \times 10^{51}$ ergs s$^{-1}$).  For larger $B_{\nu}$, $\tau_{{\rm dyn}}$ decreases to a fraction of the PNS rotation period due to magneto-centrifugal acceleration around the $T=0.5$ MeV radius; the field required to minimize $\tau_{{\rm dyn}}$ is approximately $B_{{\rm cf}}$ (eq.~[\ref{bcrit}]).  For very rapid rotation ($P \lesssim$ 2-3 ms), centrifugally-enhanced mass loss reduces the Alfv\'{e}n radius, which in turn decreases the effectiveness of magneto-centrifugal acceleration and increases $\tau_{{\rm dyn}}$.  The large reduction in $\tau_{{\rm dyn}}$ in strongly magnetized, rapidly rotating PNS winds provides conditions more favorable for successful $r$-process nucleosynthesis (see Fig.~\ref{plot:svsomega}).  \\
\label{plot:tdynplot}}
\end{figure}

HWQ present numerical calculations delineating the regions of ($S^{a},\tau_{{\rm dyn}},Y_{e}^{a}$) parameter space necessary for nucleosynthesis to reach the third abundance peak, assuming an adiabatic cooling model.  In general, they find that the condition for successful $r$-process takes the approximate functional form
\be  
S^{a} \gtrsim (\tau_{{\rm dyn}})^{1/3}\mathcal{F},
\label{hwqthreshold}
\ee
where $\mathcal{F}$ is a function of only $Y_{e}^{a}$ (see HWQ eqs. [20a,b] for analytic approximations to these results).  Hence, at fixed neutron abundance (i.e., fixed $Y_{e}^{a}$) a successful $r$-process is favored for large values of the ratio $(S^{a})^{3}/\tau_{{\rm dyn}}$.  However, for a wide range of reasonable PNS properties, detailed NRNM wind calculations show that this ratio falls short of that required to reach the third peak (QW; Otsuki et al.~2000; Wanajo et al.~2001; T01).  Given that magnetar birth is relatively common and that rapid rotation may be its key ingredient, we have quantified the effects that magnetar-strength fields and rapid rotation have on the PNS wind properties that determine whether third peak $r$-process is successful.

Figure \ref{plot:svsomega} shows $(S^{a})^{3}/\tau_{{\rm dyn}}$ from our magnetized wind solutions at $L_{\bar{\nu}_{e},51} = 8$ and $B_{\nu}$ = $10^{13},\,10^{14},\,2.5\times 10^{14}$, and $10^{15}$ G as a function of the PNS rotation period; also shown is $(S^{a})^{3}/\tau_{{\rm dyn}}$ for $B_{\nu} = 2.5\times 10^{14}$ G and  $L_{\bar{\nu}_{e},51} = 3.5$ (dashed line).  To put our results in context, we plot the entropy threshold given by HWQ (their Table 5) at the lowest $\tau_{{\rm dyn}}$ they consider ($\approx 5$ ms; choosing such a low $\tau_{{\rm dyn}}$ will be justified below) with a dashed line for several $Y_{e}^{a}$.  We plot several $Y_{e}^{a}$ because the entropy required for successful $r$-process depends sensitively on $Y_{e}^{a}$, but a modest change to $Y_{e}^{a}$ in our wind calculations would not significantly alter $S^{a}$ or $\tau_{{\rm dyn}}$ from those obtained with $Y_{e}^{a}=0.4$.  Figure \ref{plot:svsomega} shows that the presence of a magnetar-strength field and mildly rapid rotation moves the critical wind parameters almost an order-of-magnitude closer to successful third peak $r$-process in the space of $(S^{a})^{3}/\tau_{{\rm dyn}}.$\footnote{This conclusion disagrees with those of Nagataki $\&$ Kohri (2001), who also considered the effects of rotation and magnetic fields on PNS winds; however, these authors were unable to consider fields larger than $\sim 10^{11}$ G because of the complicated critical point topology they encountered in computing more highly-magnetized wind solutions.} We examine the reasons for this result below. 

In rapidly rotating, strongly-magnetized PNS winds, centrifugal-flinging pushes matter quickly through the heating region, which reduces $S^{a}$.  Our calculations find that, for $B_{\nu} \gtrsim B_{{\rm cf}}$ (eq.~[\ref{bcrit}]), $S^{a}$ is approximately given by
\be 
S^{a} \simeq S_{{\rm NRNM}}^{a}\exp[-\Omega/\Omega_{S}], 
\label{sa}
\ee 
where $\Omega_{S} \simeq 3500(L_{\bar{\nu}_{e},51}/8)^{0.15}$ s$^{-1}$ and $S_{{\rm NRNM}}^{a} \simeq 70(L_{\bar{\nu}_{e},51}/8)^{-0.2}$ is the asymptotic entropy for NRNM winds in units of $k_{B}$/baryon.

At face value, the exponential decrease in entropy implied by equation (\ref{sa}) appears to stifle the chances for successful third peak $r$-process in rapidly rotating, strongly-magnetized PNS winds.  However, because such PNS winds are magnetically driven, $\tau_{{\rm dyn}}$ will also decrease with increasing $\Omega$; hence, success for the $r$-process depends on the competition between changes in $\tau_{{\rm dyn}}$ and $S^{a}$.  As Figure \ref{plot:temperature} illustrates, in magnetically-driven PNS winds the $T = 0.5$ MeV radius ($R_{0.5{\rm MeV}}$) is generally outside the sonic point and the heating region (interior to which $S^{a}$ is set); therefore, while periods $\lesssim$ 2-3 ms are required to significantly affect the dynamics in the subsonic heating region and alter $\dot{M}$ or $S^{a}$ (the ``$\dot{M}$ Enhanced'' region of Fig.~\ref{plot:regimes}), $\tau_{{\rm dyn}}$ is significantly reduced for more modest rotation rates (the much larger ``Magnetically-Driven'' region in Fig.~\ref{plot:regimes}).  

Figure \ref{plot:tdynplot} shows $\tau_{{\rm dyn}}$ from our magnetized wind solutions at $L_{\bar{\nu}_{e},51} = 8$ and $B_{\nu}$ = $10^{13},\,10^{14},\,2.5\times 10^{14},$ and $10^{15}$ G as a function of the PNS rotation period; also shown is $\tau_{{\rm dyn}}$ for $B_{\nu} = 2.5\times 10^{14}$ G and  $L_{\bar{\nu}_{e},51} = 3.5$ (dashed line).  For rotation periods $\sim 1-10$ ms and $B_{\nu}\gtrsim 10^{15}$ G our solutions obtain $\tau_{{\rm dyn}} \approx 0.3-3$ ms, just a fraction of the rotation period, while for slow rotation $\tau_{{\rm dyn}}$ approaches a value $\approx$ 20 ms for $L_{\bar{\nu}_{e},51} = 8$, similar to that derived by QW (their eq.~[61]).  As a comparison of the $B_{\nu} = 2.5\times 10^{14}$ and $10^{15}$ G solutions in Figure \ref{plot:tdynplot} shows, the effects of magnetic fields on $\tau_{{\rm dyn}}$ saturate for sufficiently large fields.  Indeed, the monopole surface field required to minimize $\tau_{{\rm dyn}}$ is approximately $B_{{\rm cf}}$ (eq.~[\ref{bcrit}]).  Ignoring centrifugal enhancement of $\dot{M}$ and using equations (\ref{mdotnrnm}) and (\ref{bmoneff}), this corresponds to a surface dipole field $\sim 10^{15}(P/4\,{\rm ms})^{3/2}(L_{\bar{\nu}_{e},51}/8)^{1.25}$ G; hence, surface fields typical of observed Galactic magnetars are probably sufficient to minimize $\tau_{{\rm dyn}}$ at early times in a PNS's thermal evolution.  Thus, if observed Galactic magnetars were born with periods in the range 2 ms $\lesssim P \lesssim 10$ ms we conclude that there is an enhanced likelihood that $r$-process was successful in their PNS winds. 

We should caution that the comparison between our calculations and the thresholds of HWQ in Figure \ref{plot:tdynplot} was made for $\tau_{{\rm dyn}} = 5$ ms, which was the most rapid outflow HWQ considered; this is, however, almost an order of magnitude longer than the dynamical time associated with some of our solutions.  Constraints similar to HWQ at $\tau_{\rm dyn}\sim3$\,ms are obtained by Meyer \& Brown (1997).  A somewhat different threshold is obtained
by Sasaqui et al.~(2006), who emphasize a previously neglected light-element $r$-process seed production channel.  In fact, recent work by Meyer (2002) shows that the threshold for $r$-process nucleosynthesis may be modified at very short $\tau_{\rm dyn}$ compared to standard expectations (i.e., $Y_e^a$ must be less than 0.5).  In order to accurately assess whether $r$-process nucleosynthesis in proto-magnetar winds will proceed to the third abundance peak and beyond, or whether the nucleosynthesis that results from these outflows may be used to constrain the birth rate of proto-magnetars, a survey of nucleosynthesis calculations should be carried out at short $\tau_{\rm dyn}$.  In fact, the critical question of whether this modest $S^{a}$, very-low $\tau_{{\rm dyn}}$ mode of $r$-process can reproduce the seemingly universal solar abundance curve above Ba (e.g., Cowan et al.~2005) must ultimately be answered through detailed nucleosynthesis calculations on our wind solutions (such calculations are currently in progress) and by including a better treatment of $Y_{e}$ in the wind.  The effects of GR not included in our calculations will tend to increase $S^{a}$ and $Y_{e}^{a}$, on balance probably increasing the likelihood of successful $r$-process (Fuller \& Qian 1996; Cardall \& Fuller 1997).

T01 found that in NRNM winds $S^{a} \propto (\tau_{{\rm dyn}})^{0.2}$; thus, even though $S^{a} \propto L_{\bar{\nu}_{e}}^{-0.2}$ increases with time as the PNS cools, it is difficult for a NRNM wind that does not produce conditions favorable for third peak $r$-process
(eq.~[\ref{hwqthreshold}]) at early times to enter the regime for a successful $r$-process at later times.  We might expect a modification to the PNS ($S^{a},\tau_{{\rm dyn}}$) evolutionary track in the case of a proto-magnetar because the dynamical time in a
magnetically-driven PNS wind is no longer set solely by the neutrino heating.  In order to address this question, we explored how $(S^{a})^{3}/\tau_{{\rm dyn}}$ changes as the PNS cools.  As shown in Figure \ref{plot:svsomega}, from our calculations at $L_{\bar{\nu}_{e},51} = 3.5$ and $B_{\nu} = 2.5\times 10^{14}$ G we find that $(S^{a})^{3}/\tau_{{\rm dyn}}$ changes by less than a factor of 2 from those at $L_{\bar{\nu}_{e},51} = 8$ and $B_{\nu} = 2.5\times 10^{14}$ G; in particular, the peak value of $(S^{a})^{3}/\tau_{{\rm dyn}}$ at $P\sim 3$ ms remains essentially unchanged.  Thus the conditions for $r$-process in magnetically-driven PNS winds do not vary strongly with luminosity over the range we have explored.

To conclude this section, we briefly consider what constraints can be placed on $r$-process in proto-magnetar winds if they are to be capable of having a significant effect on the heavy element abundance evolution of the Galaxy.  If most SNe produce $r$-process elements, the total $r$-process-rich wind mass ejected per SN must be $\sim 10^{-6} - 10^{-5}$ M$_{\sun}$ to account for the total Galactic yield (e.g., Qian 2000); hence, because the magnetar birthrate is $\sim$ 10$\%$ of the total neutron star birthrate, at least $\sim
10^{-5}-10^{-4}$ M$_{\sun}$ of $r$-process-rich material must be ejected per magnetar birth (this number is quite uncertain because the magnetar birthrate is uncertain; see Woods $\&$ Thompson 2004).  Because $r$-process in proto-magnetar winds is only likely to be successful for $P\gtrsim 2-3$ ms, rotation does not significantly enhance mass loss from the PNS (Fig.~\ref{plot:mtot}).  As a result, the required yield of $r$-process rich material per magnetar birth is similar to our estimates for the total mass extracted in NRNM PNS winds ($\sim 10^{-4}$ M$_{\sun}$).  Thus we conclude that if proto-magnetar winds are the dominant source for Galactic r-process, the r-process probably must occur early in the PNS cooling evolution (in the first few seconds, even earlier than for normal PNSs)\footnote{Of course, if the magnetar birthrate has been underestimated and is comparable to the total NS birthrate then $r$-process could occur somewhat later in the PNS evolution.}.  The former in part justifies our concentration on high $L_{\bar{\nu}_{e}}$ in Figure \ref{plot:svsomega} and means that the question of whether magnetar birth is a significant source for Galactic $r$-process is especially sensitive to the early evolution of the magnetic field and radius of the PNS.

It is also worth noting that on the basis of a comparison between the solar $r$-process abundance pattern with meteoritic abundances of $^{129}$I and $^{182}$Hf, Qian, Vogel, \& Wasserburg (1998) (see also Wasserburg et al.~1996) suggest a diversity of $r$-process sites.  In particular, they argue for a site with high frequency (roughly the Galactic SN rate) that generates the $A \approx 195$ nuclei and a site 10 times less frequent that produces nuclei near $A\approx 135$.  In order to satisfy the observational constraints, the latter site must eject 10 times the mass of the high-frequency site, per event.  Although we have not proven that magnetars produce an $r$-process in any mass range (we save a detailed investigation for a future work) the $\sim 10\%$ birth fraction of magnetars, the characteristically larger total ejected mass (Fig.~[\ref{plot:mtot}]), and the very different thermodynamics of their winds relative to NRNM PNS winds make it tempting to associate proto-magnetar winds with the low-frequency enrichment events advocated by Qian, Vogel \& Wasserburg (1998).

\subsubsection{Wave Heating}
\label{section:waveheating}

QW show that by including a heating source in addition to neutrinos outside a few PNS radii, the wind entropy is increased, the dynamical time is reduced, and the chances for a successful $r$-process can be substantially improved.  Such an extended heating mechanism operates above the sun, where convectively excited waves are believed to heat the extended solar corona (for recent work see Cranmer $\&$ van Ballegooijen 2005).  The neutrino cooling luminosity of a PNS also drives vigorous convection during the Kelvin-Helmholtz epoch (e.g., Burrows $\&$ Lattimer 1986; Burrows, Hayes, $\&$ Fryxell 1995).  It is likely that a fraction of the convective energy flux will be deposited into outgoing waves, which will then propagate into the PNS atmosphere and deposit their energy and momentum on a length scale of order a few PNS radii.  The relative importance of hydrodynamic and MHD wave excitation likely depends on the magnetic field strength of the PNS.  Here we focus on heating by MHD waves in the magnetospheres of strongly magnetized PNSs (Suzuki \& Nagataki 2005), though hydrodynamic wave excitation may be important as well (e.g., Burrows et al.~2006a,b).  If the energy flux in MHD waves at the PNS surface is $F_{{\rm w}}(R_{\nu}) = (1/2)\rho v_{A}^{3}(\delta B/B_{r})^{2}|_{{R_{\nu}}}$, we estimate that the total wave heating $\dot{Q}_{{\rm{w}}} \approx 4\pi R_{\nu}^{2}F_{{\rm w}}(R_{\nu})$ is given by
\be 
\dot{Q}_{{\rm w}} \simeq 10^{48}(B_{15})^{3}\left(\frac{\delta B_{\nu}/B_{\nu}}{0.1}\right)^{2}{\rm ergs}\,\,{\rm s^{-1}},
\label{qdottotwave}
\ee 
where $\delta B_{\nu} \equiv \delta B(R_{\nu})$ is the amplitude of the waves excited at the PNS surface (with a density $\rho(R_{\nu}) \approx 10^{12}$ g cm$^{-3}$).  The ratio of the wave heating in equation ($\ref{qdottotwave}$) to the total neutrino heating $\dot{Q}_{\nu}= 4\pi\int_{R_{\nu}}^{\infty}r^{2}\rho\dot{q}_{\nu}dr$ in the absence of rotation can be approximated as 
\be
\frac{\dot{Q}_{{\rm w}}}{\dot{Q}_{{\rm \nu}}} \approx
0.03(B_{15})^{3}\left(\frac{\delta
B_{\nu}/B_{\nu}}{0.1}\right)^{2}\left(\frac{L_{\bar{\nu}_{e},51}}{8}\right)^{-2.8},
\label{qratio}
\ee 
where $\dot{Q}_{\nu} \approx 4.4\times 10^{49}(L_{\bar{\nu}_{e},51}/8)^{2.8}$ ergs s$^{-1}$ is taken from our NRNM calculations and agrees reasonably well with the results of T01 (their Table 1).  Equation (\ref{qratio}) illustrates that, for efficient wave excitation, wave heating at early times may become important for field strengths $\gtrsim 10^{14}-10^{15}$ G.  Equation (\ref{qratio}) also appears to suggest that wave heating will become substantially more important as the PNS cools and $L_{\bar{\nu}_{e}}$ drops; however, whether this in fact occurs is unclear because the energy flux in waves and the surface amplitude $\delta B_{\nu}$ will likely decrease with $L_{\bar{\nu}_{e}}$ as the convective flux decreases.

In order to quantify the effects that wave heating have on the entropy and dynamical time in rotating PNS winds, we consider a concrete model in which we add Alfv\'{e}n wave pressure and heating to our solutions.  In the entropy equation (eq. [\ref{entropyeq}]), this leads to an additional source term of the form 
\be \dot{q}_{{\rm w}}(r) = \frac{v_{A}}{v_{r}+v_{A}}\frac{F_{{\rm w}}(R_{\nu})}{\chi R_{\nu}\rho}\left(\frac{R_{\nu}}{r}\right)^{2}\exp\left[\frac{-(r-R_{\nu})}{\chi R_{\nu}}\right],
\label{qdotwave}
\ee 
where the factor $v_{A}/(v_{r}+v_{A}) < 1$ accounts for the work done by the Alfv\'{e}n waves.  Equation (\ref{qdotwave}) concentrates the total wave heating ($\dot{Q}_{{\rm w}} = 4\pi\int_{R_{\nu}}^{\infty}r^{2}\rho\dot{q}_{{\rm w}}dr$) on a radial length scale $\approx \chi R_{\nu}$.  Radially propagating Alfv\'{e}n waves also exert a pressure on the fluid (${\rm P}_{{\rm w}}$), which contributes a term to the right hand side of the radial momentum equation (eq.~[\ref{radmomentum}]) of the form (Lamers \& Cassinelli 1999; Suzuki \& Nagataki 2005) 
\be 
-\frac{1}{\rho}\frac{d{\rm P}_{{\rm w}}}{dr} = \frac{\dot{q}_{{\rm w}}}{2(v_{r}+v_{A})} - \frac{(\delta B)^{2}}{32 \pi \rho}\frac{3v_{r} + v_{A}}{v_{r} + v_{A}}\frac{1}{\rho}\frac{d\rho}{dr}, \ee where \be (\delta B)^{2} = \frac{8\pi\chi R_{\nu}\rho\dot{q}_{{\rm w}}}{v_{r}+v_{A}}. 
\ee 
We consider a variety of dissipation lengths ($R_{\nu}\chi$) and surface wave amplitudes ($\delta B_{\nu}$) and assess the effects of Alfv\'{e}n wave heating on our wind solutions.  For reasons discussed at the end of $\S$\ref{section:rprocess}, we are primarily interested in the effect that wave heating has on $r$-process at early times; thus, we concentrate on wave heating applied to high luminosity solutions.

Our wave heating calculations are summarized in Table \ref{table:waveheatingtable}.  Motivated by the $r$-process threshold of HWQ (eq.~[\ref{hwqthreshold}]), we quantify the improvement towards a successful $r$-process through an ``improvement factor'': 
\be
I_{{\rm w}}(\dot{Q}_{{\rm w}}, \chi) \equiv \frac{[S_{0.5{\rm MeV}}^{3}/\tau_{{\rm dyn}}]_{\dot{Q}_{{\rm w}}}}{[S_{0.5{\rm MeV}}^{3}/\tau_{{\rm dyn}}]_{\dot{Q}_{{\rm w}}=0}},
\label{improvement}
\ee 
where $S_{0.5{\rm MeV}}$ is the entropy at the radius where $T=0.5$ MeV.  Figure \ref{plot:svsomega} shows that for $L_{\bar{\nu}_{e},51} = 8, B_{\nu} = 10^{15}$ G, and $P = 3$ ms (corresponding to the peak in Fig. \ref{plot:svsomega}) an improvement factor $I_{{\rm w}} \gtrsim 6$ is required to exceed the HWQ threshold for third-peak $r$-process for $Y_{e}^{a} \approx 0.4$, while for a NRNM solution at $L_{\bar{\nu_{e}},51} = 8$ successful $r$-process requires $I_{{\rm w}} \gtrsim 25$.  We should note, however, that HWQ assumed adiabatic expansion through the $\alpha$-process temperature range ($T \approx 0.5-0.2$ MeV), while for large $\chi$ wave heating is important at these radii and the expansion is not adiabatic.  This may modify the $I_{{\rm w}}$ required for successful $r$-process.

For $P \simeq 3(130)$ ms, $B = 10^{15}$ G, and $L_{\bar{\nu}_{e},51} = 8$, we find that $I_{{\rm w}} \gtrsim 6(25)$ requires an amplitude $\delta B_{\nu}/B_{\nu} \gtrsim 0.5-0.7$ and a dissipation length $\chi \sim 3$.\footnote{For very small $\chi$, the wave heating effectively acts as an increase to the neutrino luminosity, which decreases the asymptotic entropy $S^{a}$ (recall that -- absent wave heating -- the entropy decreases with increasing luminosity).  For large $\chi$, on the other hand, the heating is concentrated outside the $T=0.5$ MeV radius and, while $S^{a}$ increases substantially, the entropy at $T=0.5$ MeV remains relatively unaffected.}  This wave amplitude corresponds to a total wave heating $\gtrsim 4\times 10^{49}$ ergs s$^{-1}$, which is comparable to the total neutrino heating for this solution and is $\sim 10^{-3}$ of the neutrino luminosity at early times.  The fact that a comparable wave energy is required in the slow and rapidly rotating limits appears somewhat coincidental.  For the non-rotating solutions, the wave heating only reduces the dynamical time to $\sim 2$ ms, never reaching the regime of $\tau_{{\rm dyn}}\lesssim$ 1 ms obtained in rotating PNS winds.  The entropy increase at $T=0.5$ MeV is, however, larger in the non-rotating case because the $T=0.5$ MeV radius is at larger radii (and thus $S_{0.5{\rm MeV}}$ is closer to $S^{a}$).  Although the actual wave dissipation mechanism in the PNS atmosphere is uncertain, we note that for initial amplitudes of $\delta B_{\nu}/B_{\nu} \sim 0.5$, conservation of action implies that the waves are nonlinear with $\delta B/B_{r}\sim 1$ at $\sim 3-4 R_{\nu}$, similar to the dissipation lengths ($\chi$) that are optimal for $r$-process.  Also note that for more rapidly rotating solutions ($P \approx 1$ ms), significantly more wave heating is required for successful r-process: even $\delta B_\nu \approx B_\nu$ is insufficient at $B \approx 10^{15}$ G.  This is because the total neutrino heating itself increases at $P \sim 1$ ms ($\rho$ is larger in the neutrino heating region due to magneto-centrifugal support), so a given amount of wave heating has less of an effect on the solution.  

For lower neutrino luminosities, smaller $\dot{Q}_{{\rm w}}$ can lead to successful $r$-process.  For example, at $L_{\bar{\nu}_{e},51} = 3.5$ and $B = 2.5\times 10^{14}$ G, we find that, for $P$ = 3(130) ms, $\dot{Q}_{{\rm w}} \gtrsim 10^{49}(5\times 10^{48}$) ergs s$^{-1}$ and $\chi \sim 3$ is required to eclipse the HWQ $r$-process threshold for $Y_{e}^{a} = 0.4$.\footnote{Suzuki \& Nagataki (2005) found that for non-rotating solutions with Alfv\'{e}n wave heating, $r$-process was successful for $\dot{Q}_{{\rm w}} \simeq 2\times 10^{48}$ ergs s$^{-1}$ and $\chi \sim 5-10$ at $L_{\bar{\nu}_{e},51} \sim 1$.  Our results are similar to theirs although their required wave heating is somewhat larger than an extrapolation of our results would suggest.  However, they used different mean neutrino energies than we have assumed and obtain $\dot{M}$ $\sim 2\times 10^{-6}$ M$_{\sun}$ s$^{-1}$, roughly 3 times greater than we predict from equation (\ref{mdotnrnm}) at $L_{\bar{\nu}_{e},51} = 1$.  This larger mass-loading may explain why their calculations required somewhat more wave heating to enter the regime of successful $r$-process.}  The required wave heating in this case is again comparable to the total neutrino heating, but is only $\sim 6\times 10^{-4}(3 \times 10^{-4})$ of the neutrino luminosity of the PNS.  At even lower $L_\nu$, a yet smaller fraction of the neutrino luminosity in wave heating would be capable of yielding successful $r$-process.  However, as discussed in the previous section, constraints on the r-process rich material required per magnetar-birth imply that the r-process must be successful at relatively high neutrino luminosities.  We thus conclude that wave heating leads to successful $r$-process if $\gtrsim 10^{-4}-10^{-3}$ of the PNS's neutrino luminosity emerges in wave power at early times in the Kelvin Helmholtz epoch, and that this required level of wave heating is essentially independent of magnetic field strength and rotation rate (for $P \gtrsim 2$ ms).  

Although the mechanism and formula we describe here are appropriate to proto-magnetars with large magnetic fields, it is possible that hydrodynamic (as opposed to MHD) wave heating is important and generic to normal PNS birth.  In this case, waves may be generated by convective motions as the PNS cools, or via global modes of the PNS similar to those observed by Burrows et al.~(2006a,b), that persist into the cooling epoch.   As in the MHD case, a fraction of the total neutrino luminosity ($10^{-4}-10^{-3}$) must emerge in wave power over $\tau_{\rm KH}$ to produce conditions suitable for the $r$-process.  Because such a mechanism may operate generically (and not just in proto-magnetars) the requirement of an early-time $r$-process highlighted above for the proto-magnetars and in \S4.3 is somewhat alleviated.

\subsection{Additional Applications}
\label{section:AIC}

While most magnetar formation probably results from the core collapse of a massive star, magnetars may also form through the accretion-induced collapse (AIC) of a white dwarf (Nomoto et al.~1979; Usov 1992; Woosley $\&$ Baron 1992).  Rapid rotation will automatically accompany AIC due to the accretion of mass and angular momentum prior to collapse, and a strong field may accompany the final stages of the PNS's contraction, amplified through either magnetic flux-freezing of the progenitor white dwarf's field or via dynamo action (Duncan \& Thompson 1992).  The properties of PNSs formed following AIC would thus be very similar to those of the proto-magnetars that we have considered in this paper and the resulting proto-magnetar wind would be accurately modeled using our calculations.  AIC may thus give rise to a LGRB with properties similar to those considered in $\S$\ref{section:LGRB}.  However, LGRBs from AIC will not produce significant amounts of Ni (Woosley \& Baron 1992; Dessart et al.~2006) and hence will not be associated with a simultaneous hyper-energetic Type-Ic SN, as in, e.g. GRB 980425/SN1998bw or GRB 030329/2003dh (Galama et al.~1998; Hjorth et al.~2003; Stanek et al.~2003).  Instead, AIC should be associated with a class of SN-less LGRBs like GRB 060505 and 060614 (Fynbo et al.~2006; Gal-Yam et al.~2006; Della Valle et al.~2006).  A prediction of this model is that these LGRBs should be associated with both relatively old (few Gyr) and relatively young ($\sim$100\,Myr) stellar populations.  If some SN-less LGRBs are found to be associated with old stellar populations, it would strongly support the AIC interpretation.

Although we have focused on magnetized PNS evolution in this paper, the physical conditions we have considered are quite similar to those expected in a neutrino-cooled accretion disk surrounding a newly-formed black hole in the ``collapsar'' model for LGRBs (Woosley 1993; Paczynski 1998; MacFadyen $\&$ Woosley 1999).  For concreteness, consider the properties of a thin accretion disk with an accretion rate $\dot{M}_{{\rm acc}} = 0.2 M_{\sun}$ s$^{-1}$ and viscosity parameter $\alpha = 0.01$ at a fiducial radius of $\approx 100$ km, or approximately 10 Schwarzschild radii above a non-rotating black hole of mass $3 M_{\sun}$.  According to the calculations of Chen $\&$ Beloborodov (2006), at these radii (where $P \approx 10$ ms) the disk is optically thick to neutrinos, the surface temperature of the disk is very similar to that of a PNS neutrinosphere ($\approx 2-3$ MeV), and the total neutrino luminosity of the disk is $L_{\nu} \simeq 0.04\dot{M}_{{\rm acc}}c^{2} \approx 1.4\times 10^{52}$ ergs s$^{-1}$ (their Fig. 18), comparable to that of a PNS at early times (note that at lower accretion rates the disk will be optically thin to neutrinos).  If this disk were threaded with a large-scale poloidal field with a strength corresponding to that expected for MRI turbulence, $B \approx 10^{14}-10^{15}$ G (plasma $\beta \approx 10-100$), the physical conditions would indeed resemble those in proto-magnetar winds.  Thus neutrino-magneto-centrifugal driving may be important in setting the mass-loading and energy loss rate in outflows from collapsar disks.  Indeed, Levinson (2006) has calculated the mass loading of neutrino-driven outflows in general relativistic MHD for conditions anticipated in collapsar disks, finding qualitatively similar results to those discussed in this paper for PNS winds.

As a final context in which our calculations may be relevant, we note that a magnetized accretion flow like that considered above for collapsars or a short-lived, rapidly rotating proto-magnetar may be formed following the merger of a NS-NS binary (Paczynski 1986; Eichler et al.~1989; Rosswog \& Liebend{\"o}rfer 2003).  Our calculations would also describe magnetized outflow from these objects, but additional work is needed to explore this application in more detail.

\section{Conclusions}

\label{section:conclusions}

We have solved the one-dimensional non-relativistic neutrino-heated MHD wind problem in order to study the effects that magnetic fields and rotation have on PNS wind evolution following the launch of the SN shock but prior to the end of the Kelvin-Helmholtz cooling epoch. Figure $\ref{plot:regimes}$ summarizes the physical regimes of PNS winds in the presence of rotation and magnetic fields.  We map our monopole solutions onto the axisymmetric, relativistic dipole calculations of B06, thus taking into account the effects that neutrino-driven mass loss have on the fraction of open magnetic flux and the PNS spin-down rate.  Our primary conclusions are as follows:

\begin{itemize}
\item{We identify three types of PNSs based on the dominant character of their energy loss during the Kelvin-Helmholtz epoch (see Figure \ref{plot:etot}): 
\begin{itemize}
\item[(1)]{ For slow rotation and low magnetic field strengths $(B_{14}^{{\rm dip}} \lesssim 10^{-2}P_{{\rm ms}}^{2}$, where $B^{{\rm
dip}} = 10^{14} B_{14}^{{\rm dip}}$ G is the surface dipole field strength of the PNS and $P = 1 P_{{\rm ms}}$ ms is its birth period),
a neutrino-heated, thermally-driven wind dominates energy and mass loss from the PNS; most radio pulsars were probably this type at
birth.}
\item[(2)]{ For larger rotation rates and field strengths $(10^{-2}P_{{\rm ms}}^{2} \lesssim B_{14}^{{\rm dip}} \lesssim 2P_{{\rm ms}}^{2}\exp[2P_{{\rm ms}}^{-2}])$, a non-relativistic, magnetically-driven wind dominates during the Kelvin-Helmholtz epoch.  Most observed Galactic magnetars ($\sim 10\%$ the birthrate of radio pulsars) were probably of this type at birth.  For $B \sim 10^{15}$ G and $P \lesssim$ 2 ms, greater than $10^{51}$ ergs can be lost to a non-relativistic, magnetically-driven outflow during $\tau_{KH}$, and, over a broad range of initial spin period, the energy extracted is many times larger than from a slowly rotating PNS (see Fig.~\ref{plot:etot}).  The outflow from this type of PNS is likely to be collimated by magnetic stresses and the asymmetric injection of energy may be sufficient to generate an anisotropy in the morphology of the SNe remnant.}
\item[(3)]{For rapid rotation and field strengths somewhat larger than those observed from Galactic magnetars $(B_{14}^{{\rm dip}} \gtrsim 2P_{{\rm ms}}^{2}\exp[2P_{{\rm ms}}^{-2}])$, a relativistic, magnetically-driven wind dominates energy loss during the Kelvin-Helmholtz epoch.  Although the birthrate of PNSs of this type is probably small (if they are produced at all), they may be capable of producing hyper-energetic SNe and long-duration gamma-ray bursts.}
\end{itemize}
}

\item{For $P \approx$ 3(1) ms and $B_{\nu}^{{\rm dip}} \gtrsim 10^{15}$ G the total mass loss during the Kelvin-Helmholtz epoch is enhanced by a factor of $\approx$ 2($10^{2}$) relative to a non-rotating PNS (Fig. \ref{plot:mtot}).}

\item{For initial PNS spin periods of $P \approx $ 1 ms and magnetic field strengths of $B_{\nu}^{{\rm dip}} \approx 3\times 10^{15}-10^{16}$ G, we find that $\gtrsim 10^{52}$ ergs of rotational energy is extracted on a timescale of $10-40$ seconds and that the magnetization $\sigma$ of the outflow is $\sim 0.1-1000$.  The energy, luminosity, timescale, and mass-loading of the late-time outflow are all consistent with those required to explain long duration gamma-ray bursts (assuming efficient dissipation of magnetic energy into kinetic energy at large radii; e.g., Drenkhahn \& Spruit 2002).  In addition, outflows from such PNSs have the property that the PNS rotational energy is extracted with a roughly uniform distribution in log($\sigma$) over the timescale $\tau_{{\rm KH}}$.  For these modest $\sigma$ winds, especially at early times, energy loss from the PNS is enhanced relative to pure force-free spindown because of additional open magnetic flux (see the discussion in $\S$\ref{section:evolution}).  Thus, a significant portion of the PNS rotational energy can be extracted in just a few seconds following the launch of the SN shock, perhaps sufficiently rapidly to increase the nucleosynthetic yield of the SN (e.g., $^{56}$Ni).}

\item{Winds from PNSs with $B_{\nu}^{{\rm dip}}\gtrsim 10^{15}$ G and $P \approx 2-10$ ms produce conditions almost an order-of-magnitude more favorable for third-peak $r$-process nucleosynthesis in the space of $(S^{a})^{3}/\tau_{{\rm dyn}}$ than do winds from more slowly-rotating, less-magnetized PNSs (see Fig. \ref{plot:svsomega}).  For these rotation rates, the asymptotic entropy is similar to that of a non-rotating, non-magnetized wind, while the dynamical time is significantly reduced by magnetic acceleration (see Fig. \ref{plot:tdynplot}).  The very different thermodynamic properties of rapidly rotating proto-magnetar winds (relative to non-rotating PNS winds) may contribute to the inferred diversity of $r$-process sites (e.g., Qian et al.~1998).}

\item{Heating by outgoing hydrodynamic or MHD waves may be important in PNS winds.  We find that if $\gtrsim 10^{-4}-10^{-3}$ of the PNS's neutrino luminosity emerges in wave power at early times in the Kelvin-Helmholtz epoch, then r-process is successful in PNS winds.  This conclusion is relatively independent of magnetic field strength and rotation period for $P \gtrsim$ 2 ms.}
\end{itemize}
\acknowledgements We thank Niccolo Bucciantini and Jon Arons for helpful discussions.  We also thank Martin White and Marc Davis for computational time on their Beowulf cluster.  T.A.T. thanks the Aspen Center for Physics, where this work germinated, for its hospitality.  EQ and BDM were  supported in part by NSF grant AST 0206006, NASA grants NAG5-12043 and NNG05GO22H, an Alfred P. Sloan Fellowship, the David and Lucile Packard Foundation, and a NASA GSRP Fellowship to BDM.  Wind profiles are available upon request from BDM.

\newpage

\begin{table}
\begin{scriptsize}
\begin{center}
\caption{PNS Wind Properties at $L_{\bar{\nu}_{e}} = 8\times 10^{51}$ ergs s$^{-1}$, $R_{\nu}$ = 10 km, and $M$ = 1.4 M$_{\sun}$
\label{table:solutiontable}}

\begin{tabular}{lccccccccccc}
\hline \hline

\\

\multicolumn{1}{c}{ $B_{\nu}$ } &
\multicolumn{1}{c}{ $\Omega$} &
\multicolumn{1}{c}{ $P$ } &
\multicolumn{1}{c}{ $\dot{M}$ } &
\multicolumn{1}{c}{ \hspace{1cm}$\sigma$\hspace{1cm} }&
\multicolumn{1}{c}{ $\tau_{J}$ }&
\multicolumn{1}{c}{ $S$ \tablenotemark{a}}&
\multicolumn{1}{c}{ $\tau_{{\rm dyn}}$ \tablenotemark{b}}&
\multicolumn{1}{c}{ $R_{{\rm s}}$ \tablenotemark{c} }&
\multicolumn{1}{c}{ $R_{A}$ }&
\multicolumn{1}{c}{ $\dot{E}$ \tablenotemark{d} }&
\multicolumn{1}{c}{ $\eta$ \tablenotemark{e} }\\

\multicolumn{1}{c}{(G)} &
\multicolumn{1}{c}{(s$^{-1}$)} &
\multicolumn{1}{c}{(ms)} &
\multicolumn{1}{c}{$({\rm M_{\sun}\,s^{-1}})$} &
\multicolumn{1}{c}{} &
\multicolumn{1}{c}{(s)} &
\multicolumn{1}{c}{(k$_{{\rm B}}$/baryon)} &
\multicolumn{1}{c}{(ms)} &
\multicolumn{1}{c}{(km)} &
\multicolumn{1}{c}{(km)} &
\multicolumn{1}{c}{($10^{51}$ ergs s$^{-1}$)} &
\multicolumn{1}{c}{}\\

\\
\hline
\\
$10^{15}$   &  6000  &  1.0  &  $1.2\times 10^{-2}$  &  0.055  &  4.8  &  13.7  &  0.52  &  19  &  31  &  4.7  &  1.80  \\ 
$10^{15}$   &  4000  &  1.6  &  $1.0\times 10^{-3}$  &  0.30  &  12.6  &  24.1  &  0.48  &  23  &  67  &  1.2  &  1.18  \\ 
$10^{15}$   &  2000  &  3.1  &  $2.8\times 10^{-4}$  &  0.26  &  12.9  &  40.6  &  1.1  &  35  &  120  &  0.31  &  1.14  \\ 
$10^{15}$   &  1000  &  6.3   &  $1.9\times 10^{-4}$  &  0.099  &  9.3  &  52.2  &  2.5  &  55  &  180 &  0.11  &  1.11   \\ 
$10^{15}$   &  500   &  13  &  $1.6\times 10^{-4}$  &  0.029  &  6.2  &  59.5 &  5.4  &  87  &  240  &  $4.0\times 10^{-2}$  &  1.14  \\ 
$10^{15}$   &  250   &  25  &  $1.5\times 10^{-4}$  &  8.0$\times 10^{-3}$  &  4.0  &  64.5  &  9.8  &  140  &  310  &  1.5$\times 10^{-2}$  &  1.18  \\ 
$10^{15}$   &  50   &  130  &  $1.4\times 10^{-4}$  &  3.4$\times 10^{-4}$  &  1.6  &  69.0  &  21  &  420  &  500  &  $1.7\times 10^{-3}$ &  1.00  \\ 
\\
\hline
\\
$10^{14}$   &  6000  &  1.0  &  $3.3\times 10^{-3}$  &  $2.0\times 10^{-3}$  &  39  &  20.2  &  5.5  &  68  &  21  &  0.13  &  7.9  \\ 
$10^{14}$   &  4000  &  1.6  &  $6.6\times 10^{-4}$  &  $4.5\times 10^{-3}$  &  101  &  29.9  &  2.8  &  48  &  29  &  0.046  &  3.8  \\ 
$10^{14}$   &  2000  &  3.1  &  $3.8\times 10^{-4}$  &  $3.0\times 10^{-3}$  &  116  &  46.2  &  3.7  &  76  &  44  &  0.013  &  2.9  \\ 
$10^{14}$   &  1000  &  6.3  &  $2.4\times 10^{-4}$  &  $1.1\times 10^{-3}$  &  88  &  57.8  &  7.1  &  150  &  61  &  4.6$\times 10^{-3}$  &  2.7  \\ 
$10^{14}$   &  500   &  13  &  $1.7\times 10^{-4}$  &  $3.1\times 10^{-4}$  &  64  &  64.8  &  13  &  310  &  77  &  1.8$\times 10^{-3}$  &  2.5  \\ 
$10^{14}$   &  250   &  25  &  $1.5\times 10^{-4}$  &  $8.2\times 10^{-5}$  &  52  &  68.0  &  19  &  520  &  88  &  7.9$\times 10^{-4}$  &  1.7  \\ 
$10^{14}$   &  50   &  130  &  $1.4\times 10^{-4}$  &  $3.4\times 10^{-6}$  &  45  &  69.8  &  24  &  730  &  95  &  3.6$\times 10^{-4}$  &  0.17  \\ 
\\
\hline
\\

$10^{13}$   &  6000  &  1.0  &  $6.0\times 10^{-4}$  &  1.1$\times 10^{-4}$  &  350  &  39.1  &  29  &  530  &  16  &  2.7$\times 10^{-3}$  &  43  \\
$10^{13}$   &  4000  &  1.6  &  $2.5\times 10^{-4}$  &  1.2$\times 10^{-4}$  &  670  &  52.6  &  21  &  510  &  18  &  1.4$\times 10^{-3}$  &  19  \\ 
$10^{13}$   &  2000  &  3.1  &  $1.6\times 10^{-4}$  &  4.6$\times 10^{-5}$  &  850  &  64.3  &  22  &  640  &  20  &  6.5$\times 10^{-4}$  &  13 \\  
$10^{13}$   &  1000  &  6.3   &  $1.4\times 10^{-4}$  &  1.3$\times 10^{-5}$  &  880  &  68.3  &  23  &  710  &  21  &  4.2$\times 10^{-4}$  &  3.0 \\ 
$10^{13}$   &  500   &  13  &  $1.4\times 10^{-4}$  &  3.3$\times 10^{-6}$  &  880  &  69.7  &  23  &  740  &  21  &  3.7$\times 10^{-4}$  &  0.85  \\ 
$10^{13}$   &  250   &  25  &  $1.4\times 10^{-4}$  &  8.5$\times 10^{-7}$  &  880  &  69.7  &  24  &  740  &  22  &  3.5$\times 10^{-4}$  &  0.23  \\ 
$10^{13}$   &  50   &  130  &  $1.4\times 10^{-4}$  &  3.4$\times 10^{-8}$  &  880  &  69.9  &  24  &  750  &  22  &  3.4$\times 10^{-4}$  &  0.0093  \\ 
\\
\hline
\hline
\end{tabular}
\end{center}
\end{scriptsize}

\tablenotetext{a}{The asymptotic wind entropy.}
\tablenotetext{b}{The dynamical time evaluated at $T=0.5$ MeV (see eq.~[\ref{taudef}]).}
\tablenotetext{c}{The radius of the adiabatic sonic point.  For large $B_{\nu}$ and $\Omega$ the slow point and adiabatic sonic point are very close to each other, while for low $B_{\nu}$ and $\Omega$ they approach the Alfv\'{e}n and fast magnetosonic radii, respectively.}
\tablenotetext{d}{The asymptotic wind power.}
\tablenotetext{e}{The ratio of the rotational power lost by the PNS to the asymptotic wind power (see eq.~[\ref{eta1}]).}

\end{table}

\clearpage

\begin{table}
\begin{scriptsize}
\begin{center}
\caption{PNS Wind Properties at $B_{\nu} = 2.5\times 10^{14}$ G, $R_{\nu}$ = 10 km, and $M$ = 1.4 M$_{\sun}$
\label{table:solutiontable2}}

\begin{tabular}{lccccccccccc}
\hline \hline

\\

\multicolumn{1}{c}{ $L_{\bar{\nu}_{e},51}$ \tablenotemark{a}} &
\multicolumn{1}{c}{ $\Omega$} &
\multicolumn{1}{c}{ $P$ } &
\multicolumn{1}{c}{ $\dot{M}$ } &
\multicolumn{1}{c}{ \hspace{1cm}$\sigma$\hspace{1cm} }&
\multicolumn{1}{c}{ $\tau_{J}$ }&
\multicolumn{1}{c}{ $S$ \tablenotemark{b}}&
\multicolumn{1}{c}{ $\tau_{{\rm dyn}}$ \tablenotemark{c}}&
\multicolumn{1}{c}{ $R_{{\rm s}}$ \tablenotemark{d} }&
\multicolumn{1}{c}{ $R_{A}$ }&
\multicolumn{1}{c}{ $\dot{E}$ \tablenotemark{e} }&
\multicolumn{1}{c}{ $\eta$ \tablenotemark{f} }\\

\multicolumn{1}{c}{} &
\multicolumn{1}{c}{(s$^{-1}$)} &
\multicolumn{1}{c}{(ms)} &
\multicolumn{1}{c}{$({\rm M_{\sun}\,s^{-1}})$} &
\multicolumn{1}{c}{} &
\multicolumn{1}{c}{(s)} &
\multicolumn{1}{c}{(k$_{{\rm B}}$/baryon)} &
\multicolumn{1}{c}{(ms)} &
\multicolumn{1}{c}{(km)} &
\multicolumn{1}{c}{(km)} &
\multicolumn{1}{c}{($10^{51}$ ergs s$^{-1}$)} &
\multicolumn{1}{c}{}\\

\\
\hline
\\
\,\,$8$ &  6000  &  1.0  &  $6.2\times 10^{-3}$  &  0.0067  &  17.0  &  16.5  &  2.2  &  31  &  23  &  0.57  &  4.1  \\ 
\,\,$8$ &  4000  &  1.6  &  $8.3\times 10^{-4}$  &  0.022  &  50.7  &  26.2  &  1.1  &  28  &  37  &  0.17  &  2.0  \\ 
\,\,$8$ &  2000  &  3.1  &  $2.6\times 10^{-4}$  &  0.017  &  56.3  &  42.8  &  1.8  &  43  &  61  &  0.047  &  1.7  \\ 
\,\,$8$ &  1000  &  6.3   &  $1.8\times 10^{-4}$  &  6.4$\times 10^{-3}$  &  40.4  &  54.3  &  3.8  &  75  &  87 &  0.016  &  1.7   \\ 
\,\,$8$ &  500   &  13  &  $1.5\times 10^{-4}$  &  1.9$\times 10^{-4}$  &  26.7  &  61.8  &  7.8  &  140  &  117  &  $6.1\times 10^{-3}$  &  1.7  \\ 
\,\,$8$ &  250   &  25  &  $1.4\times 10^{-4}$  &  5.0$\times 10^{-4}$  &  18.1  &  66.4  &  14  &  270  &  150  &  2.4$\times 10^{-3}$  &  1.6  \\ 
 \,\,$8$ &  50   &  130  &  $1.4\times 10^{-4}$  &  2.1$\times 10^{-5}$  &  11.6  &  69.5  &  23  &  680  &  190  &  $4.6\times 10^{-4}$ &  0.52  \\ 
\\
\hline
\\
$3.5$   &  6000  &  1.0  &  $1.6\times 10^{-3}$  &  0.027  &  47  &  14.3  &  0.66  &  20  &  28  &  0.37  &  2.3  \\ 
$3.5$   &  4000  &  1.6  &  $1.1\times 10^{-4}$  &  0.17  &  150  &  27.6  &  0.81  &  23  &  57  &  0.091  &  1.3  \\ 
$3.5$   &  2000  &  3.1  &  $3.6\times 10^{-5}$  &  0.13  &  152  &  48.0  &  2.5  &  36  &  100  &  0.025  &  1.2  \\ 
$3.5$   &  1000  &  6.3  &  $2.4\times 10^{-5}$  &  0.048  &  108  &  62.0  &  6.7  &  58  &  150  &  8.5$\times 10^{-3}$  &  1.2  \\ 
$3.5$   &  500   &  13  &  $2.1\times 10^{-5}$  &  0.014  &  69  &  71.2  &  14.3  &  94  &  200  &  3.2$\times 10^{-3}$  &  1.3  \\ 
$3.5$   &  250   &  25  &  $1.9\times 10^{-5}$  &  $3.8\times 10^{-3}$  &  43  &  77.1  &  26  &  160  &  260  &  1.2$\times 10^{-3}$  &  1.4  \\ 
$3.5$   &  50   &  130  &  $1.8\times 10^{-5}$  &  $1.6\times 10^{-4}$  &  15.8  &  83.1  &  53  &  550  &  450  &  1.1$\times 10^{-4}$  &  1.5  \\ 
\\
\hline
\hline
\end{tabular}
\end{center}
\end{scriptsize}

\tablenotetext{a}{The anti-electron neutrino luminosity in units of $10^{51}$ ergs s$^{-1}$.}
\tablenotetext{b}{The asymptotic wind entropy.}
\tablenotetext{c}{The dynamical time evaluated at $T=0.5$ MeV (see eq.~[\ref{taudef}]).}
\tablenotetext{d}{The radius of the adiabatic sonic point.  For large $B_{\nu}$ and $\Omega$ the slow point and adiabatic sonic point are very close to each other, while for low $B_{\nu}$ and $\Omega$ they approach the Alfv\'{e}n and fast magnetosonic radii, respectively.}
\tablenotetext{e}{The asymptotic wind power.}
\tablenotetext{f}{The ratio of the rotational power lost by the PNS to the asymptotic wind power (see eq.~[\ref{eta1}]).}

\end{table}

\newpage

\begin{table}
\begin{scriptsize}
\begin{center}
\vspace{0.05 in}\caption{PNS Wind Properties with Wave Heating at $L_{\bar{\nu}_{e},51}$ = 8 and $B_{\nu} = 10^{15}$ G}
\label{table:waveheatingtable}

\begin{tabular}{lccccccccccc}
\hline \hline

\\

\multicolumn{1}{c}{$\Omega$} &
\multicolumn{1}{c}{$P$ } &
\multicolumn{1}{c}{$(\delta B/B)|_{R_{\nu}}$ \tablenotemark{a}} &
\multicolumn{1}{c}{$\chi$} &
\multicolumn{1}{c}{$\dot{Q}_{\nu}$ \tablenotemark{b}} &
\multicolumn{1}{c}{$\dot{Q}_{{\rm w}}$ \tablenotemark{c}} &
\multicolumn{1}{c}{$\dot{M}$ }&
\multicolumn{1}{c}{$\tau_{{\rm dyn}}$ \tablenotemark{d}}&
\multicolumn{1}{c}{$S_{0.5{\rm MeV}}$ \tablenotemark{e}}&
\multicolumn{1}{c}{$R_{{\rm 0.5\,MeV}}$ \tablenotemark{f} }&
\multicolumn{1}{c}{$I_{{\rm w}}$ \tablenotemark{g}}\\

\multicolumn{1}{c}{(s$^{-1}$)} &
\multicolumn{1}{c}{(ms)} &
\multicolumn{1}{c}{} &
\multicolumn{1}{c}{} &
\multicolumn{1}{c}{($10^{50}$ ergs s$^{-1}$)} &
\multicolumn{1}{c}{($10^{50}$ ergs s$^{-1}$)} &
\multicolumn{1}{c}{$({\rm M_{\sun}\,s^{-1}})$} &
\multicolumn{1}{c}{(ms)} &
\multicolumn{1}{c}{(k$_{{\rm B}}$/baryon)} &
\multicolumn{1}{c}{(km)} &
\multicolumn{1}{c}{}\\

\\
\hline
\\
 50   & 130 & -   & - & 0.44 & 0    & 1.4$\times 10^{-4}$ & 21.0 & 69.0 & 73.2 & 1 \\
 50   & 130 & 0.2 & 3 & 0.39 & 0.054  & 1.6$\times 10^{-4}$ & 11.4 & 75.0 & 65.0 & 2.37 \\
 50   & 130 & 0.7 & 1 & 0.38 & 0.65 & 3.4$\times 10^{-4}$ & 4.2 & 92.5 & 62.5  & 12.0 \\ 
 50   & 130 & 0.7 & 3 & 0.44 & 0.37 & 2.9$\times 10^{-4}$ & 1.9 & 119 & 51.2 & 55 \\
 50   & 130 & 0.7 & 10 & 0.44 & 0.48 & 2.4$\times 10^{-4}$ & 2.0 & 88.5 & 44.9 & 22.2 \\
 2000   & 3.1 & - & - & 0.53 & 0    & 2.8$\times 10^{-4}$ & 1.07 & 40.6 & 33.4 & 1 \\
 2000   & 3.1 & 0.7 & 0.3 & 0.30 & 0.65 & 4.8$\times 10^{-4}$ & 0.96  & 44.1 & 37.4 & 1.4 \\
 2000   & 3.1 & 0.7 & 1 & 0.54 & 0.64 & 5.0$\times 10^{-4}$ & 0.92 & 68.9 & 41.4 & 5.7 \\
 2000   & 3.1 & 0.7 & 3 & 0.56 & 0.57 & 4.3$\times 10^{-4}$ & 1.02 & 78.6 & 40.3 & 7.6 \\
 2000   & 3.1 & 0.7 & 10 & 0.60 & 0.42 & 3.7$\times 10^{-4}$ & 1.00 & 55.3 & 35.2 & 2.7 \\
 2000   & 3.1 & 1.0 & 3 & 0.63 & 1.76 & 5.6$\times 10^{-4}$ & 0.99 & 109 & 46.8 & 21.2 \\
 6000 & 1.0 & -   & - & 5.57 & 0    & 1.2$\times 10^{-2}$ & 0.52 & 13.7 & 30.8 & 1 \\
 6000 & 1.0 & 0.7 & 3 & 6.53 & 0.36 & 1.3$\times 10^{-2}$ & 0.74 & 14.5 & 35.9 & 0.83 \\
 6000 & 1.0 & 1.0 & 1 & 6.04 & 0.74 & 1.4$\times 10^{-2}$ & 0.79 & 15.6 & 41.8 & 0.96  \\
 6000 & 1.0 & 1.0 & 3 & 5.91 & 0.74 & 1.3$\times 10^{-3}$ & 0.94 & 15.4 & 41.1 & 0.80 \\ 

\\
\hline
\hline
\end{tabular}
\end{center}
\end{scriptsize}

\tablenotetext{a}{The fractional wave amplitude at the PNS surface.}
\tablenotetext{b}{Total neutrino heating rate.}
\tablenotetext{c}{Total wave heating rate.}
\tablenotetext{d}{The dynamical time evaluated at $T=0.5$ MeV (see eq.~[\ref{taudef}]).}
\tablenotetext{e}{The wind entropy at $T=0.5$ MeV.  Note that $S_{0.5{\rm MeV}} \ne S^{a}$ because, in general, wave heating extends outside the $T=0.5$ MeV radius.}
\tablenotetext{f}{The radius where $T=0.5$ MeV.}
\tablenotetext{g}{The ratio of $(S_{0.5{\rm MeV}})^{3}/\tau_{{\rm dyn}}$ with wave heating to without wave heating (eq.~[\ref{improvement}]).}

\end{table}

\end{document}